\documentclass[10pt]{article}
\usepackage{amsmath,amsfonts,vmargin}
\usepackage{amsbsy}
\usepackage{verbatim}
\usepackage{setspace}
\usepackage{rotating}
\usepackage{array}
\usepackage{graphicx}
\usepackage{pdflscape}
\usepackage{graphicx,epsf}
\usepackage[T1]{fontenc}
\usepackage[utf8]{inputenc}
\usepackage{tabularx,ragged2e,booktabs}
\usepackage{color,soul}
\usepackage{xcolor}

\newcommand{\bz}{{\bf z}}

\newcommand{\bgam}{\mbox{\boldmath{$\gamma$}}}

\newcommand{\bomg}{\mbox{\boldmath{$\Omega$}}}

\newcommand{\bZ}{{\bf Z}}
\newcommand{\bA}{{\bf A}}
\newcommand{\bI}{{\bf I}}

\newcommand{\Cov}{\mbox{Cov}}

\newcommand{\mc}{\multicolumn}
\newcommand{\real}{\mathbb{R}}

\newcommand{\bth}{{\mbox{\boldmath{$\theta$}}}}

\newcommand{\bPs}{{\mbox{\boldmath{$\Psi$}}}}

\newcommand{\bph}{{\mbox{\boldmath{$\phi$}}}}
\newcommand{\bhph}{{\mbox{\boldmath{$\hat{\phi}$}}}}
\newcommand{\bphi}{{\mbox{\boldmath{$\varphi$}}}}

\newcommand{\bhtht}{\mbox{\boldmath{$\hat{\theta}$}}}

\numberwithin{equation}{section}
\usepackage{xcolor,soul}

\title{Correcting for Measurement Error in Segmented Cox Model}
\author{Sarit Agami\\
Department of Economics\\
Hebrew University, Mount Scopus, Jerusalem, Israel\\
email:sarit.agami@mail.huji.ac.il}

\begin{document}
\maketitle


\begin{abstract}
Measurement error in the covariate of main interest (e.g. the exposure variable, or the risk factor) is common in epidemiologic and health studies. It can effect the relative risk estimator or other types of coefficients derived from the fitted regression model. In order to perform a measurement error analysis, one needs information about the error structure. Two sources of validation data are an internal subset of the main data, and external or independent study. For the both sources, the true covariate is measured (that is, without error), or alternatively, its surrogate, which is error-prone covariate, is measured several times (repeated measures). This paper compares the precision in estimation via the different validation sources in the Cox model with a changepoint in the main covariate, using the bias correction methods RC and RR. The theoretical properties under each validation source is presented. In a simulation study it is found that the best validation source in terms of smaller mean square error and narrower confidence interval is the internal validation with measure of the true covariate in a common disease case, and the external validation with repeated measures of the surrogate for a rare disease case. In addition, it is found that addressing the correlation between the true covariate and its surrogate, and the value of the changepoint, is needed, especially in the rare disease case.
\end{abstract}


\maketitle

\section{Introduction \label{section intro}}
Measurement error is common in many empirical studies. This error refers to the discrepancy between the true value and the measured value of a covariate (Thomas et al., 1993). It occurs when a covariate cannot be observed exactly (usually due to instrument or sampling error), and therefore we observe a surrogate covariate instead of measuring the true covariate. The surrogate measures the true covariate with some error.
For example, in clinical trials, biomarkers such as blood pressure (Carroll
et al., 1995) and CD4 counts (Tsiatis et al., 1995), are subject to measurement error; fat intake in
nutritional studies relies largely on self-reported measures of dietary intake, and therefore is often measured with error (Carroll et al., 1995, Keogh 2013).
Ignoring such error can substantially degrade the quality of inference or even yield erroneous
results (see, for example, Carroll et al., 2006). Therefore, ways to correct the impact of this error are needed. There are two broad strategies to correct for measurement error (Buonaccorsi, 2010):
The first is to estimate the required corrections for the model parameters from a validation study, independent of the main study under consideration, that is, an external study; The second is to measure the required corrections within the main study, that is, an internal study. In both external and internal studies, the sample may include the true covariate, or alternatively, repeated measures of the error-prone covariate. That is, we can distinguish between four sources of validation data:
\noindent 1. \textit{External} \textit{Validation }data (EV) -- the main study includes measures of the surrogate covariate only, but there exists a previous study, external to the main study, which includes measures of surrogate covariate and the true covariate. Then the connection between these two covariates can be estimated using this external data.
\\
\noindent 2. \textit{Internal} \textit{Validation} data (IV) -- the true covariate is available on a subset of the main study.
\\
\noindent 3. \textit{Repeated measures, External} (RM external) --  the true covariate is not observed for any subject, but repeated measures of surrogate covariate are available on subjects that are outside the main study.
\\
\noindent 4. \textit{Repeated measures, Internal} (RM internal) --  the true covariate is not observed for any subject, but repeated measures of surrogate covariate are available on a subset of the main study.
\\
\indent Internal validation study has the capability of direct examination of the error structure. With external validation study, there may be doubt as to whether the populations characterized by the validation and main study samples are comparable in the sense that the measurement error model is equivalent or "transportable" between the populations. Based on validity concerns alone, internal validation study may has the greatest advantage. However, this neglects the important issue of the costs of obtaining the true exposures, which may be considerably larger than those for a more readily available surrogate.
\\
\indent Validation for measurement error correction had been examined widely in the literature. For example,  Lyles et al. (2007) studied the efficiency of combining internal and external validation data in a case-control setting; Wong et al. (1999) studied the design of validation study for estimation the correction factor in univariate and bivariate situation; Muoka et al. (2019) proposed a method that adjusts for measurement error in three correlated exposures in the absence of internal validation study; Siddique at al. (2019) proposed a statistical framework for correcting for measurement error in longitudinal self-reported dietary data using an external validation study; Spiegelman and Gray (1991) studied cost-efficient study designs for binary response data with Gaussian covariate measurement error; Thurigen et al. (2000) presented a review of methods and their applicability in case-control studies for Measurement error correction using validation data.
\\
\indent The aim of the current paper is to study the validation for measurement error correction in the segmented Cox model, which was introduced by Agami et al. (2020). Specifically, the aim is to examine the influence of the validation type on the precision of the model parameters estimation. Agami et al. (2020) studied several bias correction methods, and examined their performance under repeated measures from an external reliability design. In this paper, some of these bias correction methods are considered, and the theoretical properties under each source validation is developed for the considered bias correction methods. The paper is organized as follows: Section 2 presents the setting, the segmented Cox model, and the considered bias correction methods. Section 3 presents the theoretical properties under each validation data. Section 4 presents a simulation study comparing the model parameters estimation under the various bias correction methods based on the various validation data. The comparison is done in terms of point estimation and confidence intervals.
Section 5 presents a brief summary.

%
%

\section{The Model}
Let $X(t)$ denotes the covariate of main interest (which can depend on time), and let $\bZ(t) \in \real^p$ denotes the vector of additional covariates. Define $a_+ = \max(a,0)$. The Cox model describes the hazard as a function of the covariates $X(t)$ and $\bZ(t)$, and the baseline hazard function $\lambda_0(t)$. Once the influence of $X(t)$ on the hazard changes at some level of $X(t)$, the segmented Cox model is of the form:
\begin{equation}
\lambda(t|x(t),\bz(t)) = \lambda_0(t) \exp(\bgam^T \bz(t) + \beta x(t) + \omega (x(t)-\tau)_+).
\label{CP1}
\end{equation}
Assuming a known changepoint $\tau$, the unknown parameters are $\bth = (\bgam ^{T} ,\beta ,\omega )$.
When it is impossible or too expansive to observe the covariate $X(t)$, we observe instead a surrogate covariate $W(t)$, which measures the covariate $X(t)$ with error. The assumptions throughout this paper are:
\\
(i) The error is additive, that is, $W(t) = X(t) + U$, where $U$ is a random variable such that $E(U|X(t))=0$ (this is the classical error model).
\\
(ii) $\bZ(t) \in \real^p$ is error-free
\\
(iii) The conditional distribution of $X(t)$ given $\bZ(t)=\bz$ is $N(\mu_x(\bz),\sigma_x^2)$, and the $U(t)$'s are i.i.d.\
$N(0,\sigma_u^2)$, independent of the $X(t)$'s and the $\bZ(t)$'s.
\\
(iv) $\mu_x(\bz)$ is of the form $\mu_x(\bz) = \alpha_0 + \alpha_1 \bz$.
\\
The observations are on $n$ independent individuals.
For a given individual $i$, $\tilde{T}_i$ denotes the time of entry into the study, $T_i^\circ$ denotes the survival time, and $C_i$ denotes the time of right censoring. The event indicator is $\delta_i = I(T_i^\circ \leq C_i)$, and the at-risk indicator is denoted by $Y_i(t)= I(\tilde{T}_i\leq t \leq T_i^\circ)$.
The maximum possible follow-up time is denoted by $t^{*}$.
Let us write the relative risk as $r(x,\bz,\bth) = \exp(\bgam^T \bz + \beta x + \omega (x - \tau)_+)$. If $X(t)$ was known, the standard Cox log partial likelihood is given by
\begin{align*}
l_{p} (\bth) & = \sum _{i=1}^{n} \delta _{i}  [ \log r(X_i(t),\bZ_i(t),\bth)
-\log \sum _{j=1}^{n}Y_{j}(T_{i}) r(X_i(t),\bZ_i(t),\bth)] \\
& = \sum _{i=1}^{n} \delta _{i} [ (\bgam^T \bZ_i(t) + \beta X_i(t) + \omega (X_i(t) - \tau)_+)
- \log \sum _{j=1}^{n}Y_{j}(T_{i}) r(X_i(t),\bZ_i(t),\bth) ].
\label{lik}
\end{align*}

Generally, the bias correction methods for Cox regression analysis with covariate error, as were studied in Agami et al. (2020), involve replacing $r(x,\bz,\bth)$
with some substitute, are as follows:
\\
\noindent A. Regression Calibration (RC) Methods

\noindent A1.  Simple RC Method (RC1): $X_{i}(t)$ is replaced with
$\mu(W_i(t),\bZ_i(t))=E(X_{i}(t) |W_{i}(t), \bZ_i(t))$.

\noindent A2.  Improved RC Method (RC2): $X_{i}(t)$ is replaced with $E( X_{i}(t)|W_{i}(t),
\bZ_i(t))$ and $(X_{i}(t) -\tau )_{+} $ is replaced with $E( (X_{i}(t) -\tau )_{+}
|W_{i}(t), \bZ_i(t) )$.
\\
\noindent B. Induced Relative Risk (RR) Methods

\noindent B1. Original RR Method (RR1): This is an extension of the method proposed by Prentice (1982) to threshold models; $r(x,\bz,\bth) = E[\exp(\bgam^T \bz + \beta x + \omega (x - \tau)_+)|W(t),\bZ(t)]$.

\noindent B2. Modified RR Method (RR2): This is a bootstrap bias-correction procedure based on the weighted bootstrap algorithm as in Kosorok and Song (2007).

\section{Estimation}
\subsection{Background}
The type of the validation data is involved in the log partial likelihood, and therefore has influence on the point estimators of the model parameters, and their corresponded confidence intervals as well.
\noindent The RC and RR estimators are obtained as the solution to $U(t^{*},\bth)=0$, where $U(t^{*},\bth)$ is the score function based on the above log partial likelihood.
Denote by $\hat{\bth}$ the resulting estimator. In addition, let $u^{(g)}(t^{*},\bth)$ be the limit of the score function $U(t^{*},\bth)$. and denote by $\bth ^{*}$ the solution of $u(t^{*},\bth)=0$. As was showed in Agami et al. (2020), $n^{1/2} (\hat{\bth }-\bth ^{*} )$ converges in distribution
to a mean-zero multivariate normal distribution whose covariance matrix can be consistently estimated by
$$
\Omega (t,\, \hat{\bth },\, g)=
[n^{-1} I(t,\, \hat{\bth},\, g)]^{-1} \hat{A}(t,\, \hat{\bth },\, g) [n^{-1}I(t,\, \hat{\bth },\, g)]^{-1}.
$$
The explicit expression for each quantity depends on the specific bias correction method.
This asymptotic covariance matrix of $\bhtht$ involves the nuisance parameters vector $\bph :=\left(\mu _{x} ,\sigma _{x}^{2} \, ,\sigma _{u}^{2} \, \right)$.
$\bph$ is usually unknown and need to be estimated, and the error in estimating $\bph$ must be accounted for in the asymptotic covariance matrix of $\bhtht$.

\subsection{Point Estimation and Confidence Interval}
As was mentioned in Section 1, the error estimation can be done by one of four sources of data: EV, IV, RM external, RM internal. Bellow, a description of the estimation procedure and the theoretical properties under each validation data is presented.
\\
Notation: Denote by $\bph ^{*} $ the true value of $\bph $. Also, denote by $\bhtht_{ns}$ the estimator of $\bth$ that involves the estimated nuisance parameters, and  by $\bI_{ns}$ the minus second derivative of the log likelihood with the nuisance parameters replaced by their estimates. Let $\dot{U}_{\bph} $ denotes the first derivative of $U$ respect to $\bph$.
\subsubsection{External Validation data (EV)}
In this setting, the main study includes measures of $W(t)$ only, but there exists a previous study, external to the main study, which includes measures of $X(t)$ and $W(t)$. Suppose this external data includes information on $m$ independent subjects. In this case, the knowledge of $X(t)$ is entered in the estimation of the nuisance parameters $\bph$ only, and therefore influence the log-likelihood through $r(X_i(t),\bZ_i(t),\bth)$.
\noindent The measurement error can be calculated as $U =W(t) -X(t)$ for the subjects with both $X(t)$ and $W(t)$ known. The estimated $\bph$, $\bhph$, may be obtained as a solution of the estimation equations of the form $\sum _{i=1}^{m}\bPs _{i} (\bph) =0$, where
\begin{align*}
\bPs _{i} (\bph)=\begin{bmatrix}
\bar{X}-\mu _{x}\\
{(X_{i} -\bar{X})^{2} -\sigma _{x}^{2} \frac{(m-1)}{m} }\\
{U _{i}^{2} -\sigma _{u }^{2} \frac{(m-1)}{m} }
\end{bmatrix}.
\end{align*}

\noindent \textbf{Proposition 1}.\textbf{ \underbar{}}

\noindent \textbf{ (i) Consistency}: $\bph ^{*} $  is the solution of  $\sum _{i=1}^{m}E(\bPs _{i} (\bph )) =0$. Then, $\hat{\bph }{\mathop{\to }\limits^{p}} \bph ^{*} $.

\noindent \textbf{(ii) Corrected Covariance Matrix:}

\noindent The corrected covariance matrix is
\begin{align*}
\bomg _{corr} (t^{*} ,\bhtht_{ns} ,\bhtht,g)=n^{-1} \bI_{ns} (t^{*} ,\, \bhtht_{ns} ,\, g)^{-1} \bA_{corr} (t^{*} ,\, \bhtht_{ns} ,\hat{\bph},\, g)n^{-1} \bI_{ns}(t^{*} ,\, \bhtht_{ns} ,\, g)^{-1}
\end{align*}
where
\begin{align*}
\bA_{corr} (t^{*} ,\bth ^{*} ,\bph ^{*} ,g)=\bA(t^{*} ,\bth ^{*} ,\bph ^{*} ,g)+\dot{U}_{\bph }(t^{*} ,\bth ^{*} ,\bph ^{*} ,g)Cov(\hat{\bph})\dot{U}_{\bph}(t^{*} ,\bth ^{*} ,\bph ^{*} ,g)^{T}.
\end{align*}
The proof is presented in Appendix A.1.
\subsubsection{Internal Validation Data (IV)}
Here, the main study includes measures of $W(t)$ for all the $n$ subjects, but there are $m$ independent subjects that have measures of $X(t)$ and $W(t)$. In this case, the knowledge of $X(t)$ is entered in the estimation of the nuisance parameters $\bph$, but also through the observations included in the log partial likelihood. Hence we use the true $X(t)$ where available, and otherwise we use its surrogate. That is, the log-likelihood is of the form:
\begin{align*}
l_{p} (\bth) & = \sum _{i=1}^{n} \delta _{i}  [ \log r(\tilde{X}_i(t),\bZ_i(t),\bth)
-\log \sum _{j=1}^{n}Y_{j}(T_{i}) r(\tilde{X}_i(t),\bZ_i(t),\bth)]
\end{align*}
where $\tilde{X}$ is a vector of length $n$ for which $m<n$ includes the information of real $X(t)$ and the other $n-m$ the relevant surrogate.
Since $X(t)$ is available for some subjects, therefore, the consistency of $\hat{\bph}{\mathop{\to }\limits^{p}} \phi ^{*} $ is the same as previously in the case of EV setting. Regarding the corrected covariance matrix, although $\hat{\bph }$ and $U(t^{*} ,\bth ^{*} ,\bph ^{*} ,g)$ involve data from the same sample, they are nonetheless asymptotically independent. This holds using the argument in Zucker and Speigelman (2008) that the contribution to $U(t^{*} ,\bth ^{*} ,\bph ^{*} ,g)$ made by each individual in the study has asymptotically expectation zero conditional on the true covariate value, and hence it follows from an iterated expectation argument that $U(t^{*} ,\bth ^{*} ,\bph ^{*} ,g)$ and $\hat{\bph}$ are asymptotically uncorrelated. For more details, see Zucker and Spiegelman (2008), Section 3. It then follows that the corrected covariance matrix in the case of IV setting is the same as in the previous case of EV setting.

\subsubsection{RM External}
In repeated measures setting, the vector $\hat{\bph}$ is obtained as a solution of the estimation equations of the form $\sum _{i=1}^{m}\bPs _{i} (\bph) =0$, where $m$ is a sample of independent individuals and
\begin{align*}
\bPs _{i} (\bph)=\begin{bmatrix}
k_{i} (\bar{W}_{i\bullet } -\mu _{x})\\
{k_{i} (\bar{W}_{i\bullet } -\mu _{x} )^{2} -\sigma _{u}^{2} \frac{(m-1)}{m}-[k_{i} -\frac{k_{i}^{2} }{\sum _{i=1}^{m}k_{i}  } ]\sigma _{x}^{2}}\\
{\sum _{j=1}^{k_{i} }(W_{ij} -\bar{W}_{i\bullet } )^{2} -(k_{i} -1)\sigma _{u}^{2}}
\end{bmatrix},
\end{align*}

\noindent with $k_{i}$ replications for each subject $i$, and $\bar{W}_{i\bullet}$ is the average of $W_i$ over the $k_{i}$ replications.

\noindent \textbf{Proposition 2}.\textbf{(i) Consistency}: $\bph ^{*} $ is the solution of  $\sum _{i=1}^{m}E(\bPs _{i} (\bph )) =0$. Then, $\hat{\bph}{\mathop{\to }\limits^{p}} \bph ^{*} $.

\noindent \textbf{(ii) Corrected Covariance Matrix:}

\noindent The corrected covariance matrix is
 \begin{align*}
\bomg _{corr} (t^{*} ,\hat{\bth}_{ns} ,\hat{\bph},g)=n^{-1} \bI_{ns} (t^{*} ,\, \hat{\bth}_{ns} ,\, g)^{-1} \bA_{corr} (t^{*} ,\, \hat{\bth}_{ns} ,\hat{\bph},\, g)n^{-1} \bI_{ns}(t^{*} ,\, \hat{\bth}_{ns} ,\, g)^{-1}
 \end{align*}
where
 \begin{align*}
\bA_{corr}(t^{*} ,\bth ^{*} ,\bph ^{*} ,g)=\bA(t^{*} ,\bth ^{*} ,\bph ^{*} ,g)+\dot{U}_{\bph} (t^{*} ,\bth ^{*} ,\bph ^{*} ,g)Cov(\hat{\bph})\dot{U}_{\bph} (t^{*} ,\bth ^{*} ,\bph ^{*} ,g)^{T} ,
 \end{align*}
$\dot{U}_{\bph}$ denotes the first derivative of $U$ respect to $\bph$.

\noindent \textbf{Proof of Proposition 2. }

\noindent The proof  is similar to the proof  in the previous case of EV or IV.

The corrected covariance matrix here is the same as in the previous cases of EV or IV settings.

\subsubsection{RM Internal}
\noindent Here, the consistency of $\hat{\bph}{\mathop{\to }\limits^{p}} \bph ^{*} $ is the same as in the previous case of RM external. Regarding the corrected covariance matrix, we have the following proposition:

\noindent \textbf{ Proposition 3}.\textbf{ Corrected Covariance Matrix:} Denote by $R$ the set of individuals in the replicated measures sample. Define
$\Phi =Cov(\frac{1}{m} \sum _{i\in R}H_{i} (t^{*} ,\bth ,\bph ,g),\sqrt{m} \, \dot{\bPs }_{\bph } (\bph ^{*}))$
where $\dot{\bPs}_{\bph} $ denotes the first derivative of $\bPs$ respect to $\bph$.

\noindent Then, the corrected covariance matrix is
\begin{align*}
\bomg _{corr} (t^{*} ,\hat{\bth }_{ns} ,\hat{\bph },g)=n^{-1} \bI_{ns} (t^{*} ,\, \hat{\bth }_{ns} ,\, g)^{-1} \bA_{corr} (t^{*} ,\, \hat{\bth }_{ns} ,\hat{\bph },\, g)n^{-1} \bI_{ns} (t^{*} ,\, \hat{\bth}_{ns} ,\, g)^{-1}
\end{align*}
where
$\bA_{corr} (t^{*} ,\bth ^{*} ,\bph ^{*} ,g)$
\begin{align*}
=\bA(t^{*} ,\bth ^{*} ,\bph ^{*} ,g)+\dot{U}_{\bph} (t^{*} ,\bth ^{*} ,\bph ^{*} ,g)Cov(\hat{\bph })\dot{U}_{\bph}(t^{*} ,\bth ^{*} ,\bph ^{*} ,g)^{T} -\Phi \ddot{\bPs }_{\bph } (\hat{\bph })^{-1} \dot{U}_{\bph }(t^{*} ,\bth ^{*} ,\bph ^{*} ,g)^{T} ,
\end{align*}
$\dot{U}_{\bph} $ denotes the first derivative of $U$ respect to $\bph$, and $\ddot{\bPs}_{\bph} $ denotes the second

\noindent derivative of $\bPs$ respect to $\bph$. The proof is presented in Appendix A.2.

\section{Simulation study}
\subsection{Design}
This section presents a simulation study comparing the four validation data for the bias correction methods RC and RR, under several scenarios.
The setting of the simulation is the same as in Agami et al. (2020), and detailed below. The results are summarized in Table 1 - Table 2 below and are presented in details in Appendix B. The following labels for the validation types are used: (i) EV\_X, external validation having measures of the real $X$ (ii) IV\_X, internal validation having measures of the real $X$ (iii) EV\_RM, external validation having repeated measures of the surrogate $W$ (iv) IV\_RM, internal validation having repeated measures of the surrogate $W$.
\subsection{Simulation Design}
The simulation included a single time-independent covariate $W$, without additional covariates $\bZ(t)$, and assumed a fixed administrative censoring at time $t^{*} =10$. The considered real values of the model were $\beta=\log(1.5)$ and $\omega=\log(2)$, that is the effect of the covariate on the hazard is initially moderate
and later becomes more pronounced.
The covariate $X$ was generated as standard normal and the event time was generated as exponential with parameter $\lambda =\lambda_{0} \exp (\beta X+\omega (X-\tau )_{+} )$.
\noindent The observed surrogate covariate value $W$ was generated using the classical measurement error model with $U\sim N(0,\sigma _{u}^{2})$. Denote by $\rho_{xw}$ the correlation between $X$ and $W$; then, the values of $\sigma_{u}^{2}$ were set to yield $\rho_{xw}= 0.8, 0.6$, or 0.4 (with the resulting $\sigma_{u}^{2}$ values being 0.56, 1.77 and 5.25). The changepoint $\tau$ was set at one of 5 points at various percentiles of the distribution of $X$:
$\Phi ^{-1} (0.1),\Phi ^{-1} (0.25),\Phi^{-1} (0.5),\Phi^{-1} (0.75),\Phi ^{-1} (0.9)$.
The performance of the estimators was examined under the common disease scenario where $n=3,000$ and the cumulative incidence was 0.5, and the rare disease scenario where $n=50,000$ and the cumulative incidence was 0.03. The value of the baseline hazard $\lambda_{0}$ was determined by the cumulative incidence and the value of the changepoint $\tau$, for each case.
\noindent The simulations results are based on 1,000 replications, and in all cases, the reported results include:
 \\
 (i) Relative bias of the median of \bhtht. That is, $(bias(\bhtht)/\bth$, when $bias(\bhtht)=median(\bhtht)-\bth$.
 \\
 (ii) Mean square error (MSE) of the median. That is, $Var(\bhtht)+bias(\bhtht)^2$.
  \\\
  (iii) Confidence interval for $\bth$, based on the asymptotic theory. This is the empirical coverage rate of nominal $95\%$ confidence interval centered at the estimated parameter value and with half-width equal to 1.96 times the estimated asymptotic standard deviation.

In order to eliminate cases of divergence, the condition of $|\bhtht|\le 4.9$ was imposed, and the results (i)-(iii) are based on replicates for which this condition was satisfied. Convergence problems arose more often when greater measurement error was considered, i.e. $\rho_{xw}=0.4$, and when the changepoint was at the lower or upper extreme of the covariate domain. The convergence percent (percentage of replications in which the estimation procedure converged) is presented Appendix B, and is denoted by "Pctgud".
The validation data (both internal and external) included a sample size of 500, with two measurements of $W$ for each subject when $X$ was unlivable in the validation data.

\subsection{Results}
Comparing the results of the estimates among the four validation types, the best validation data in terms of smaller relative bias was the IV with X for the common disease, and the EV with RM for the rare disease, in most of the scenarios. In terms of smaller MSE, the best validation data was the IV with X for the common disease. For the rare disease, the results of the MSE were depended on the correlation $\rho_{xw}$ and the value of the changepoint $\tau$; But generally for $\rho_{xw}=0.8, 0.6$ in most of the scenarios, the smaller MSE was obtained in the EV with RM, and in EV with X for $\rho_{xw}=0.4$. For more details, see the summary in Table 1. Note that convergence problems arose usually for all $\tau$ in $\rho_{xw}=0.4$ and in $\tau<0$ in $\rho_{xw}=0.6$ (for both common and rare disease). Similarly, the confidence intervals under IV with X in the common disease were narrower and closer to 0.95, and under EV with RM in the rare disease except for the RC1 which the best was IV with RM. For more details, see the summary in Table 2.

\begin{landscape}
{\bf Table 1. Best Validation Data for Point Estimation}
\begin{center}
\fontsize{12.7}{0.8}\selectfont
\begin{tabular}{m{2.09cm}|  m{1.5cm}| m{7.00cm}| m{10.0cm}}

\\
\\
\\
Disease&Corr &Relative Bias& MSE\\
\hline
\\
\\
Common$^a$&0.8&IV\_X&IV\_X \\
\\
\\
\\
\\
      &0.6&IV\_X for the all 4 validation data,&IV\_X\\
      \\
      && but IV\_RM for $\beta$ in RR methods\\
 \\
\\
\\
\\
     &0.4&IV\_X &IV\_X\\
 \\
\\
\\
\\
\\
\\
Rare$^b$&0.8&EV\_RM& Results for a given method over the 4 validation data are closed, except for $\tau=\Phi^{-1}(0.1)$.\\
\\
&&& For this $\tau$ the best is IV\_X for RC2, RR1, RR2, and EV\_RM for RC1.\\
\\
\\
\\
\\

&0.6&EV\_RM& Results for a given method over the 4 validation data are closed, except for $\tau=\Phi^{-1}(0.1), \Phi^{-1}(0.25)$.\\
\\
&&& For these two values of $\tau$, the best is EV\_RM,\\
&&& but for RR and $\tau=\Phi^{-1}(0.25)$, the best is IV\_X.\\
\\
\\
\\
\\
&0.4&EV\_RM for $\beta$. &Different results for a given method over the 4 validation data.
\\
&&No best validation data for $\omega$, but in most scenarios, EV\_RM is the best. $^c$ &
But in most scenarios, EV\_X is the best. $^c$\\
     \\

 \\
\\
\end{tabular}
\end{center}
\footnotesize{$^a$ $n=3,000$ with cumulative incidence of 0.5. $^b$ $n=50,000$ with cumulative incidence of 0.03. $^c$  Large variability of Pctgud, hence comparison is less correct.}
\end{landscape}

\begin{landscape}
{\bf Table 2. Best Validation Data for Confidence Interval}
\begin{center}
\fontsize{12.7}{0.8}\selectfont
\begin{tabular}{m{2.09cm}|  m{2.5cm}| m{10.0cm}}

\\
\\
\\
Disease&Corr &Confidence Interval\\
\hline
\\
\\
Common$^a$&0.8, 0.6, 0.4&IV\_X for RC\\
\\
\\
&& For RR1: small differences over the four validation data\\
\\
\\
&& in $\beta$,
 but the best is IV\_RM in $\omega$. \\
 \\
\\
\\
\\
\\
\\
Rare$^b$&0.8, 0.6, 0.4&IV\_RM for RC1 \\
\\
\\
&&EV\_RM for RC2\\
\\
\\
&& small differences over the four validation data for RR1\\
\\
\\
&& in both $\beta$ and $\omega$.\\

     \\

 \\
\\
\end{tabular}
\end{center}
\footnotesize{$^a$ $n=3,000$ with cumulative incidence of 0.5. $^b$ $n=50,000$ with cumulative incidence of 0.03. }
\end{landscape}

\section{Summary}
In this paper the theoretical properties of four validation types were examined, and a numerical comparison of these validation data had conducted via a simulation study. The numerical comparison was under the RC and RR bias correction methods in the setting of segmented Cox model with known changepoint. It was found that in general, the best validation data was IV with X for the common disease case. For the rare disease, which best validation type depends on the value of $\rho_{xw}$: EV with RM for $\rho_{xw}=0.8, 0.6$, and EV with X for $\rho_{xw}=0.4$.

\section*{Appendices}
\section*{Appendix A. \enspace Theoretical Proofs}
\section*{A.1. \enspace Proof of Proposition 1.}
\noindent \textbf{(i) Consistency}

\noindent The estimation equations $\sum _{i=1}^{m}\bPs _{i} (\bph ) =0$ are of the form of M-estimator. By the theory of M-estimator and if the estimation equations are unbiased, under regularity conditions, $\hat{\bph}$ is a consistent estimator of $\bph$ and its asymptotic distribution is normal.

\noindent In the considered case, it is easy to see that the estimator is unbiased and the regularity conditions are fulfilled. Therefore $\hat{\bph}$ is consistent estimator of $\bph ^{*} $ and asymptotically normal.
$\square$

\noindent \textbf{(ii) Corrected Covariance Matrix}

\noindent The following is based on Zucker and Spiegelman (2008).

\noindent Denote by $\dot{U}_{\bth} (t^{*} ,\bth ,\bph,g)$ the first derivative of $U(t^{*} ,\bth ,\bph ,g)$ respect to $\bth$, and by $\dot{U}_{\bph} (t^{*} ,\bth ,\bph ,g)$ the first derivative of $U(t^{*} ,\bth ,\bph ,g)$ respect to $\bph$.

\noindent Taylor expansion of $U(t^{*} ,\bth ,\hat{\bph},g)$ around $(\bth ^{*} ,\bph ^{*} )$ gives:
\begin{align*}
0=U(t^{*} ,\bth ,\hat{\bph },g)\dot{=}U(t^{*} ,\bth ^{*} ,\bph ^{*} ,g)+\dot{U}_{\bth} (t^{*} ,\tilde{\bth },\tilde{\bph },g)(\bth -\bth ^{*} )+\dot{U}_{\bph}(t^{*} ,\tilde{\bth},\tilde{\bph},g)(\hat{\bph}-\bph ^{*})
\end{align*}
where $\tilde{\bth}$ lies between $\bth ^{*} $ and $\bth $, and $\tilde{\bph}$ lies between $\bph ^{*} $ and $\bph$.

\noindent In particular for $\bhtht_{ns} $ we have that:
\begin{align*}
0=U(t^{*} ,\bhtht_{ns} ,\hat{\bph},g)\dot{=}U(t^{*} ,\bth ^{*} ,\bph ^{*} ,g)+\dot{U}_{\bth} (t^{*} ,\tilde{\bth },\tilde{\bph},g)(\bhtht_{ns} -\bth^{*} )+\dot{U}_{\bph}(t^{*} ,\tilde{\bth},\tilde{\bph},g)(\hat{\bph}-\bph ^{*} ).
\end{align*}
This is equivalent to:
\begin{align*}
0=U(t^{*} ,\bhtht_{ns} ,\hat{\bph},g)\dot{=}U(t^{*} ,\bth ^{*} ,\bph ,g)-\bI_{ns} (t^{*} ,\tilde{\bth },\tilde{\bph },g)(\bhtht_{ns} -\bth ^{*} )+\dot{U}_{\bph} (t^{*} ,\tilde{\bth},\tilde{\bph },g)(\hat{\phi }-\bph ^{*}).
\end{align*}
Rearranging, the last equation becomes:
\begin{align*}
-U(t^{*} ,\bth ^{*} ,\bph ^{*} ,g)\dot{=}-\bI_{ns} (t^{*} ,\tilde{\bth},\tilde{\bph },g)(\bhtht_{ns} -\bth ^{*})+\dot{U}_{\bph} (t^{*} ,\tilde{\bth},\tilde{\bph},g)(\hat{\bph}-\bph ^{*}),
\end{align*}
that is,
\begin{align*}
\, \, (\bhtht_{ns} -\bth ^{*})\dot{=}\bI_{ns}^{-1}(t^{*} ,\tilde{\bth},\tilde{\bph },g)[U(t^{*} ,\bth ^{*} ,\bph ^{*} ,g)+\dot{U}_{\bph} (t^{*} ,\tilde{\bth},\tilde{\bph },g)(\hat{\bph }-\bph ^{*} )].
\end{align*}
The normality of $(\bhtht_{ns} -\bth ^{*})$ involves two parts: the first is the convergence of $U(t^{*} ,\bth ^{*} ,\bph ^{*} ,g)$ to a mean-zero multivariate normal distribution, and the second part is checking the distribution of $(\hat{\bph }-\bph ^{*})$ and finding the limits of the matrices $\bI^{-1} (t^{*} ,\tilde{\bth},\tilde{\bph},g)$ and $\dot{U}_{\bph} (t^{*} ,\tilde{\bth},\tilde{\bph},g)$.

\noindent The proof of convergence of $U(t,\bth ^{*} ,\bph ^{*} ,g)$ to a mean-zero multivariate normal distribution is the same as in the case with no nuisance parameters, see at Agami et al. (2020).

\noindent As was mentioned in part (i) , $\hat{\bph}-\bph ^{*} \dot{\sim }N(0,Cov(\hat{\bph}))$.

\noindent In the case of the estimation equations $\sum _{i=1}^{m}\bPs _{i} (\hat{\bph }) =0$, we have by Taylor expansion:
\begin{align*}
0=\sum _{i=1}^{m}\bPs _{i} (\hat{\bph }) =\sum _{i=1}^{m}\bPs _{i}(\bph ^{*} ) +{\frac{1}{m} \sum _{i=1}^{m}\frac{\partial }{\partial \bph ^{T} } \bPs _{i}(\bph) }|_{\bph =\tilde{\bph}} (\hat{\bph}-\bph ^{*}).
\end{align*}
Applying  the law of the large numbers yields $\hat{\bph }-\bph ^{*} \dot{=}-\bA_{m}(\bph ^{*})^{-1} m^{-1} \sum _{i=1}^{m}\bPs _{i} (\bph ^{*}) $, where $\bA_{m} (\bph )=m^{-1} \sum _{i=1}^{m}E\{\frac{\partial }{\partial \bph ^{T} } \bPs _{i} (\bph )\} $.

\noindent That is, $Cov(\hat{\bph})=m^{-1} A_{m}^{-1} (\bph ^{*} )B_{m} (\bph ^{*})(A_{m}^{-1}(\bph ^{*} ))^{T} $, where $B_{m}(\bphi )=m^{-1} \sum _{i=1}^{m}Cov(\bPs _{i} (\bph )) $.

The estimators of  $A_{m} (\bph ^{*} )$ and $B_{m} (\bph ^{*} )$ are $\hat{A}_{m} =m^{-1} \sum _{i=1}^{m}\frac{\partial }{\partial \bph ^{T} } \bPs _{i} (\hat{\bph }) $

and $\hat{B}_{m} =m^{-1} \sum _{i=1}^{m}\bPs _{i} (\hat{\bph }) \bPs _{i}^{T} (\hat{\bph })$, respectively.

As in the case with no nuisance parameters (Agami et al., 2020), $n^{-1} \bI_{ns} (t^{*} ,\tilde{\bth },\tilde{\bph },g){\mathop{\to }\limits^{p}} \Sigma _{ns} (t^{*} ,\bth ^{*} ,\bph ^{*} ,g)$, where $\Sigma _{ns} (t^{*} ,\bth ,\bph ,g)$ depends on the nuisance parameters, and similarly, $\dot{U}_{\bph} (t^{*} ,\tilde{\bth},\tilde{\bph},g)$ tends to a constant that depends on $\bth ^{*} $ and $\bph ^{*} $. Therefore the asymptotic distribution of $\dot{U}_{\bph } (t^{*} ,\bth ^{*} ,\bph ^{*} ,g)(\hat{\bph }-\bph ^{*} )$ is normal.

Therefore,  $[U(t^{*} ,\bth ^{*} ,\bph ^{*} ,g)+\dot{U}_{\bph } (t^{*} ,\bth ^{*} ,\bph ^{*} ,g)(\hat{\bph }-\bph ^{*})]$ is distributed as normal with zero expectation. The covariance $Cov[U(t^{*} ,\bth ^{*} ,\bph ^{*} ,g)+\dot{U}_{\bph} (t^{*} ,\bth ^{*} ,\bph ^{*} ,g)(\hat{\bph }-\bph ^{*})]$ depends on the source data for the estimation of  $\bph$.

Since here $\hat{\bph}$ is estimated by an external data (EV design), then $\hat{\bph}$ and $U(t^{*} ,\bth ^{*} ,\bph ^{*} ,g)$ are independent. Therefore,
\begin{align*}
Cov[U(t^{*} ,\bth ^{*} ,\bph ^{*} ,g)+\dot{U}_{\bph }(t^{*} ,\bth ^{*} ,\bph ^{*} ,g)(\hat{\bph }-\bph ^{*})]\\=Cov(U(t^{*} ,\bth ^{*} ,\bph ^{*} ,g))+Cov(\dot{U}_{\bph }(t^{*} ,\bth ^{*} ,\bph ^{*} ,g)(\hat{\bph}-\bph ^{*} ))\,
\end{align*}
 Also,
 \begin{align*}
 Cov(\dot{U}_{\bph} (t^{*} ,\bth ^{*} ,\bph ^{*} ,g)(\hat{\bph }-\bph ^{*} ))=\dot{U}_{\bph } (t^{*} ,\bth ^{*} ,\bph ^{*} ,g)Cov(\hat{\bph })\dot{U}_{\bph} (t^{*} ,\bth ^{*} ,\bph ^{*} ,g)^{T}.
\end{align*}

\noindent Denote $Cov(U(t^{*} ,\bth ^{*} ,\bph ^{*} ,g))=A(t^{*} ,\bth ^{*} ,\bph ^{*} ,g)$.

\noindent Then, the corrected covariance matrix is
 \begin{align*}
\bA_{corr} (t^{*} ,\bth ^{*} ,\bph ^{*} ,g)=\bA(t^{*} ,\bth ^{*} ,\bph ^{*} ,g)+\dot{U}_{\bph } (t^{*} ,\bth ^{*} ,\bph ^{*} ,g)Cov(\hat{\bph})\dot{U}_{\bph }(t^{*} ,\bth ^{*} ,\bph ^{*} ,g)^{T},
 \end{align*}
and can be estimated by:
 \begin{align*}
\hat{\bA}_{corr} (t^{*} ,\hat{\bth }_{ns} ,\hat{\bph },g)=\hat{\bA}(t^{*} ,\hat{\bth}_{ns} ,\hat{\bph },g)+\dot{U}_{\bph} (\hat{\bth},\hat{\bph})Cov(\hat{\bph})\dot{U}_{\bph} (\hat{\bth},\hat{\bph})^{T}.
 \end{align*}
$\square$
\section*{A.2. \enspace Proof of Proposition 3.}

\noindent The proof here is similar to the proof in the case of external data, but with the additional argument which is relevant for the case of  estimation of $\phi $ by an internal data or by repeated measures:

\noindent We can express $\hat{\bph}$ by $\hat{\bph}\dot{=}-\ddot{\bPs }_{\bph } (\bph ^{*} )^{-1} \dot{\bPs }_{\bph } (\bph ^{*})$.

\noindent In addition, as already mentioned, we can express $n^{-1/2} U(t^{*} ,\bth ^{*} ,\bph ^{*} ,g)$ as a sum of $n^{-1/2} \sum h_{i} (t^{*} ,\bth ^{*} ,\bph ^{*} ,g) $ , where the estimator of $h_{i} (t^{*} ,\bth ^{*} ,\bph ^{*} ,g)$ is $H_{i} (t^{*} ,\hat{\bth },\hat{\bph },g)$. Therefore:

\noindent $A_{corr} (t^{*} ,\bth ^{*} ,\bph ^{*} ,g)=\bA(t^{*} ,\bth ^{*} ,\bph ^{*} ,g)+\dot{U}_{\bph} (t^{*} ,\bth ^{*} ,\bph ^{*} ,g)Cov(\hat{\bph })\dot{U}_{\bph } (t^{*} ,\bth ^{*} ,\bph ^{*} ,g)^{T} -\Phi \ddot{\bPs }_{\bph } (\hat{\bph })^{-1} \dot{U}_{\bph } (t^{*} ,\bth^{*} ,\bph ^{*} ,g)^{T} \,
\\
$$\Phi $ can be estimated asymptotically by $\hat{\Phi }=\frac{1}{m} \sum _{i\in R}H_{i} \left(t^{*} ,\hat{\theta }_{ns} ,\hat{\phi },g\right) \frac{\partial }{\partial \phi } \Psi _{i} \left(\hat{\phi }\right)^{T} $. Therefore,  $A_{corr} \left(t^{*} ,\theta ^{*} ,\phi ^{*} ,g\right)$ can be estimated by:

\noindent $\hat{A}_{corr} (t^{*} ,\hat{\theta }_{ns} ,\hat{\phi },g)=\hat{A}(t^{*} ,\hat{\theta }_{ns} ,\hat{\phi },g)+\dot{U}_{\phi } (\hat{\theta }_{ns} ,\hat{\phi })Cov(\hat{\phi })\dot{U}_{\phi } (\hat{\theta }_{ns} ,\hat{\phi })^{T} -\hat{\Phi }\ddot{\Psi }_{\phi } (\hat{\phi })^{-1} \dot{U}_{\phi } (t^{*} ,\hat{\theta }_{ns} ,\hat{\phi },g)^{T} \, .$

For the RC and RR methods, which involve the nuisance parameters $\bph=(\mu _{x} ,\sigma _{x}^{2} ,\sigma _{u}^{2} )$, these parameters are unknown and need to be estimated with the estimation error accounted for in the covariance matrix of the estimates. Suppose the estimation of $\bph$ is based on a sample of $m$ independent individuals, and denote by $\hat{\bph }$ the estimator of $\bph$. Then, the vector $\hat{\bph }$ is obtained as a solution of the estimation equations of the form $\sum _{i=1}^{m}\Psi _{i} (\bph ) =0$.
Denote by $\bph ^{*} $ the solution of  $\sum _{i=1}^{m}E(\Psi _{i} (\bph )) =0$.

Define
$$
\Phi = \Cov(\frac{1}{m} \sum _{i=1}^{m}{Y_{i} (t)}H_{i} (t,\bth ,\bph ,g),\sqrt{m} \, \dot{\Psi }_{\bph} (\bph ^{*} ))
$$
where $\dot{\Psi }_{\bph }$ denote the first derivative of $\Psi$ respect to $\bph$. Also, denote by ${\ddot{\Psi}_{\bph }} $ the second derivative of $\Psi$ respect to $\bph$, and by $\dot{U}^{(g)}_{\bph } $ the first derivative of $U^{(g)}$ respect to $\bph $. $\Phi $ can be estimated asymptotically by $$\hat{\Phi }=\frac{1}{m} \sum _{i=1}^{m}{Y_{i}(t)}H_{i} (t,\hat{\bth},\hat{\bph},g) \frac{\partial }{\partial \bph } \Psi _{i} (\hat{\bph })^{T}.$$ As in Zucker and Spiegelman (2008), the corrected covariance matrix is
$$
\Omega _{corr} (t,\hat{\bth},\hat{\bph },g)=n^{-1} I(t,\, \hat{\bth },\, g)^{-1} A_{corr} (t,\, \hat{\bth},\hat{\bph },\, g)n^{-1} I(t,\, \hat{\bth},\, g)^{-1}
$$
where
$\hat{A}_{corr}(t,\hat{\bth},\hat{\bph },g)$ is equal to the sum of $\hat{A}(t,\hat{\bth},\hat{\bph },g)$ and the term
$$
\dot{U}^{(g)}_{\bph } (t, \hat{\bth},\hat{\bph })\Cov(\hat{\bph })\dot{U}^{(g)}_{\bph } (t,\hat{\bth},\hat{\bph })^{T} -\hat{\Phi }\ddot{\Psi }_{\bph } (\hat{\bph })^{-1} \dot{U}^{(g)}_{\bph } (t,\hat{\bth},\hat{\bph })^{T}.
$$
$\square$

\begin{landscape}
\section*{Appendix B. \enspace Simulation Results}
\subsection*{B.1. \enspace Median Relative Bias}
{\bf Table 1. Median Relative Bias, Corelation=0.8, Common Disease$^a$, Validation Sample Size=500$^b$}
\begin{center}
\fontsize{7.5}{0.3}\selectfont
\begin{tabular}{m{1.09cm}|  m{0.74cm} m{0.74cm} m{0.74cm}  m{0.74cm}| m{0.74cm} m{0.74cm} m{0.74cm} m{0.74cm}| m{0.74cm} m{0.74cm}  m{0.74cm} m{0.74cm}| m{0.74cm} m{0.74cm}  m{0.74cm} m{0.74cm}}
\mc{17}{c}{\textbf{Finite Sample Bias$^c$ in \boldmath{$\beta$}, (\boldmath{$\beta$}, \boldmath{$\omega$})=(0.405, 0.693)}} \\
\\
\\
\\
$\tau$&&&RC1 &&&&RC2 &&&& RR1&&&& RR2\\
\hline
\\
\\
 & EV\_X &  IV\_X & EV\_RM & IV\_RM &  EV\_X &  IV\_X & EV\_RM & IV\_RM&  EV\_X &  IV\_X & EV\_RM & IV\_RM& EV\_X &  IV\_X & EV\_RM & IV\_RM\\
\\
\\
$\Phi^{-1}(0.1)$ &0.620&	0.562	&0.609&	0.700&0.901&	0.593&	0.857&	1.000&0.509&	0.481&	0.412&	0.606&-0.076&	0.132&	-0.122&	-0.126\\
 \\
 \\
  \\
  \\
  \\
  \\

  $\Phi^{-1}(0.25)$ & 0.418&	0.373&	0.393	&0.452&0.377&	0.265	&0.395&	0.449 &0.089&	0.123&	0.074&	0.118 &0.030&	0.078&	0.023&	0.068 \\
 \\
 \\
  \\
  \\
  \\
  \\
$\Phi^{-1}(0.5)$ & 0.275&	0.227&	0.275	&0.314& 0.113&	0.071	&0.114&	0.155 &-0.054&	-0.014&	-0.053&	-0.019&-0.063&	-0.021&	-0.064	&-0.031 \\
 \\
 \\
  \\
  \\
  \\
  \\

$\Phi^{-1}(0.75)$ &0.166&	0.125&	0.161	&0.200&0.012&	-0.005&	0.013&	0.046&-0.052&	-0.042&	-0.053&	-0.024 &-0.057&	-0.043&	-0.060&	-0.031 \\
 \\
 \\
  \\
  \\
  \\
  \\

$\Phi^{-1}(0.9)$ &0.072&	0.057&	0.074	&0.110&-0.013&	-0.022&	-0.015&	0.017&-0.036&	-0.036&	-0.041&	-0.007&-0.038&	-0.034&	-0.044&	-0.011  \\
\\
\\
\\
 \\
 \\
 \\
  \\
  \\
  \\

\mc{17}{c}{\textbf{Finite Sample Bias$^d$ in \boldmath{$\omega$}, (\boldmath{$\beta$}, \boldmath{$\omega$})=(0.405, 0.693)}} \\
\\
\\
\\
$\tau$&&&RC1 &&&&RC2 &&&& RR1&&&& RR2\\
\hline
\\
\\
 & EV\_X &  IV\_X & EV\_RM & IV\_RM &  EV\_X &  IV\_X & EV\_RM & IV\_RM&  EV\_X &  IV\_X & EV\_RM & IV\_RM& EV\_X &  IV\_X & EV\_RM & IV\_RM\\
\\
\\
$\Phi^{-1}(0.1)$& -0.589&	-0.525&	-0.580	&-0.600& -0.751&	-0.523	&-0.736&	-0.772& -0.519&	-0.492&	-0.469&	-0.541& -0.172&	-0.284&	-0.164&	-0.120  \\
 \\
 \\
  \\
  \\
  \\
  \\
  $\Phi^{-1}(0.25)$ & -0.502	&-0.419&	-0.487	&-0.476& -0.444&	-0.321&	-0.443&	-0.435& -0.289&	-0.275&	-0.278&	-0.258&-0.256&	-0.252&	-0.250&	-0.233
 \\
 \\
  \\
  \\
  \\
  \\
$\Phi^{-1}(0.5)$ &-0.453	&-0.369 &	-0.446 &	-0.442&-0.276&	-0.203&	-0.268	&-0.258 &-0.234&	-0.226&	-0.225&	-0.209&-0.231&	-0.216&	-0.229&	-0.211\\
 \\
 \\
  \\
  \\
  \\
  \\

$\Phi^{-1}(0.75)$ &-0.390&	-0.301&	-0.397&	-0.366& -0.197	&-0.132&	-0.198&	-0.173 &-0.273&	-0.233&	-0.272&	-0.248&-0.272&	-0.240&	-0.274&	-0.254 \\
  \\
 \\
  \\
  \\
  \\
  \\
$\Phi^{-1}(0.9)$ & -0.264&	-0.190&	-0.291&	-0.258&-0.163&	-0.099	&-0.191&	-0.144&-0.321&	-0.267	&-0.343&	-0.309&-0.324&	-0.265&	-0.347	&-0.313\\
\end{tabular}
\end{center}
\footnotesize{$^a$ $n=3,000$ with cumulative incidence of 0.5. $^b$ The estimates were obtained under unknown nuisance parameters which were estimated by a validation sample of size 500. $^c$ The values in the cells are relative bias of the median of $\beta$, i.e., (median-0.405)/0.405. $^d$ The values in the cells are relative bias of the median of $\omega$, i.e., (median-0.693)/0.693.}
\end{landscape}

\begin{landscape}
{\bf Table 2. Median Relative Bias, Corelation=0.6, Common Disease$^a$, Validation Sample Size=500$^b$}
\begin{center}
\fontsize{7.5}{0.3}\selectfont
\begin{tabular}{m{1.09cm}|  m{0.74cm} m{0.74cm} m{0.74cm}  m{0.74cm}| m{0.74cm} m{0.74cm} m{0.74cm} m{0.74cm}| m{0.74cm} m{0.74cm}  m{0.74cm} m{0.74cm}| m{0.74cm} m{0.74cm}  m{0.74cm} m{0.74cm}}
\mc{17}{c}{\textbf{Finite Sample Bias$^c$ in \boldmath{$\beta$}, (\boldmath{$\beta$}, \boldmath{$\omega$})=(0.405, 0.693)}} \\
\\
\\
\\
$\tau$&&&RC1 &&&&RC2 &&&& RR1&&&& RR2\\
\hline
\\
\\
 & EV\_X &  IV\_X & EV\_RM & IV\_RM &  EV\_X &  IV\_X & EV\_RM & IV\_RM&  EV\_X &  IV\_X & EV\_RM & IV\_RM& EV\_X &  IV\_X & EV\_RM & IV\_RM\\
\\
\\
$\Phi^{-1}(0.1)$ &0.950&	1.052&	0.711&	0.622&1.420&	0.298&	1.234&	1.485 &0.364&	0.816&	0.244&	0.084&-1.227&	0.239	&-1.168&	-1.598\\
 \\
 \\
  \\
  \\
  \\
  \\
  $\Phi^{-1}(0.25)$ & 0.592&	0.549&	0.501&	0.550&0.586&	0.159&	0.555	&0.707 &0.239&	0.255	&0.141&	0.254&-0.062	&0.177&	-0.159	&-0.166 \\
  \\
 \\
  \\
  \\
  \\
  \\
$\Phi^{-1}(0.5)$ &0.407&	0.331&	0.407	&0.493& 0.176&	0.025&	0.179&	0.282 &-0.039&	0.022&	-0.052&	0.017&-0.080&	0.000	&-0.078	&-0.025 \\
 \\
 \\
  \\
  \\
  \\
  \\
$\Phi^{-1}(0.75)$ &0.244&	0.190	&0.244	&0.325 &0.021	&-0.024&	0.018&	0.091&-0.066&	-0.038&	-0.070&	-0.011&-0.070&	-0.047&	-0.087	&-0.034 \\
 \\
 \\
  \\
  \\
  \\
  \\
$\Phi^{-1}(0.9)$ &0.103	&0.072&	0.098&	0.155& -0.025&	-0.034&	-0.023&	0.043 & -0.051	&-0.050&	-0.049&	0.013&-0.056&	-0.058&	-0.074	&-0.005  \\
\\
\\
\\
 \\
 \\
 \\
  \\
  \\
  \\
  \\
  \\

\mc{17}{c}{\textbf{Finite Sample Bias$^d$ in \boldmath{$\omega$}, (\boldmath{$\beta$}, \boldmath{$\omega$})=(0.405, 0.693)}} \\
\\
\\
\\
$\tau$&&&RC1 &&&&RC2 &&&& RR1&&&& RR2\\
\hline
\\
\\
 & EV\_X &  IV\_X & EV\_RM & IV\_RM &  EV\_X &  IV\_X & EV\_RM & IV\_RM&  EV\_X &  IV\_X & EV\_RM & IV\_RM& EV\_X &  IV\_X & EV\_RM & IV\_RM\\
\\
\\
$\Phi^{-1}(0.1)$& -0.886&	-0.864	&-0.826&	-0.674& -1.171&	-0.400	&-1.071&	-1.138 & -0.544&	-0.750	&-0.514&	-0.327& 0.377&	-0.435&	0.303	&0.609\\
 \\
 \\
  \\
  \\
  \\
  \\
  $\Phi^{-1}(0.25)$ &-0.729&	-0.617&	-0.690&	-0.660& -0.692&	-0.311&	-0.662&	-0.710&-0.509&	-0.449	&-0.460&	-0.485&-0.322&	-0.414	&-0.292	&-0.232\\
 \\
 \\
  \\
  \\
  \\
  \\
$\Phi^{-1}(0.5)$ &-0.698	&-0.530&	-0.684	&-0.680&-0.431&	-0.201&	-0.413&	-0.433 &-0.383&	-0.344	&-0.351	&-0.348 &-0.359&	-0.343	&-0.355	&-0.345 \\
 \\
 \\
  \\
  \\
  \\
  \\
$\Phi^{-1}(0.75)$ &-0.571&	-0.406&	-0.590&	-0.602&-0.309&	-0.122&	-0.289&	-0.313&-0.413&	-0.340&	-0.385	&-0.421 &-0.414&	-0.340&	-0.400&	-0.414  \\
 \\
 \\
  \\
  \\
  \\
  \\
$\Phi^{-1}(0.9)$ &-0.311	&-0.198&	-0.419&	-0.404&-0.247&	-0.070&	-0.287&	-0.298&-0.474&	-0.342	&-0.488&	-0.523 &-0.449	&-0.335&	-0.470&	-0.500 \\
\end{tabular}
\end{center}
\footnotesize{$^a$ $n=3,000$ with cumulative incidence of 0.5. $^b$ The estimates were obtained under unknown nuisance parameters which were estimated by a validation sample of size 500. $^c$ The values in the cells are relative bias of the median of $\beta$, i.e., (median-0.405)/0.405. $^d$ The values in the cells are relative bias of the median of $\omega$, i.e., (median-0.693)/0.693.}
\end{landscape}

\begin{landscape}
{\bf Table 3. Median Relative Bias, Corelation=0.4, Common Disease$^a$, Validation Sample Size=500$^b$}
\begin{center}
\fontsize{7.5}{0.3}\selectfont
\begin{tabular}{m{1.09cm}|  m{0.74cm} m{0.74cm} m{0.74cm}  m{0.74cm}| m{0.74cm} m{0.74cm} m{0.74cm} m{0.74cm}| m{0.74cm} m{0.74cm}  m{0.74cm} m{0.74cm}| m{0.74cm} m{0.74cm}  m{0.74cm} m{0.74cm}}
\mc{17}{c}{\textbf{Finite Sample Bias$^c$ in \boldmath{$\beta$}, (\boldmath{$\beta$}, \boldmath{$\omega$})=(0.405, 0.693)}} \\
\\
\\
\\
$\tau$&&&RC1 &&&&RC2 &&&& RR1&&&& RR2\\
\hline
\\
\\
 & EV\_X &  IV\_X & EV\_RM & IV\_RM &  EV\_X &  IV\_X & EV\_RM & IV\_RM&  EV\_X &  IV\_X & EV\_RM & IV\_RM& EV\_X &  IV\_X & EV\_RM & IV\_RM\\
\\
\\
$\Phi^{-1}(0.1)$ &-1.889&	1.308&	-0.613&	0.111 &1.638&	-0.348&	1.145&	1.232 &-1.718&	0.810	&-1.461&	-1.832 &-4.003&	0.174	&-3.085&	-3.534
 \\
 \\
  \\
  \\
  \\
  \\
  $\Phi^{-1}(0.25)$ &0.627	&0.718&	0.413&	0.281&0.702&	-0.071&	0.541&	0.748 &0.122	&0.309&	-0.169	&-0.280&-1.157&	0.211&	-1.096&	-1.238
 \\
 \\
  \\
  \\
  \\
  \\
$\Phi^{-1}(0.5)$ &0.484&	0.402&	0.473&	0.611& 0.223&	-0.042&	0.147&	0.323&0.006&	0.060&	-0.063&	-0.052 &-0.249&	0.028	&-0.414&	-0.372
 \\
 \\
  \\
  \\
  \\
  \\
$\Phi^{-1}(0.75)$ &0.288&	0.200&	0.256&	0.354&0.037&	-0.038&	0.009&	0.106&-0.048&	-0.027	&-0.091	&0.014& -0.102&	-0.045	& -0.291&	-0.176
 \\
 \\
  \\
  \\
  \\
  \\
$\Phi^{-1}(0.9)$ & 0.065&	0.051&	-0.105&	-0.052& -0.029&	-0.042	&-0.044	&0.047 &-0.058&	-0.055	&-0.081	&0.020&-0.093&	-0.070&	-0.246	&-0.143
\\
\\
\\
\\
 \\
 \\
 \\
  \\
  \\
  \\
  \\
  \\

\mc{17}{c}{\textbf{Finite Sample Bias$^d$ in \boldmath{$\omega$}, (\boldmath{$\beta$}, \boldmath{$\omega$})=(0.405, 0.693)}} \\
\\
\\
\\
$\tau$&&&RC1 &&&&RC2 &&&& RR1&&&& RR2\\
\hline
\\
\\
 & EV\_X &  IV\_X & EV\_RM & IV\_RM &  EV\_X &  IV\_X & EV\_RM & IV\_RM&  EV\_X &  IV\_X & EV\_RM & IV\_RM& EV\_X &  IV\_X & EV\_RM & IV\_RM\\
\\
\\
$\Phi^{-1}(0.1)$& 0.697&	-1.040&	-0.237&	-0.649& -1.388&	-0.017&	-1.175&	-1.187& 0.651&	-0.785	&0.385	&0.778& 1.967&	-0.455&	1.183&	1.669
 \\
 \\
  \\
  \\
  \\
  \\
  $\Phi^{-1}(0.25)$ &-0.851	&-0.740&	-0.812&	-0.687 & -0.866&	-0.133&	-0.806&	-0.854&-0.546	&-0.479	&-0.368&	-0.271&0.263&	-0.444&	0.024	&0.214
 \\
 \\
  \\
  \\
  \\
  \\
$\Phi^{-1}(0.5)$ &-0.840&	-0.615	&-0.825	&-0.817&-0.530	&-0.121	&-0.506	&-0.562&-0.491&	-0.393&	-0.430&	-0.442 &-0.320	&-0.391&	-0.387	&-0.337
 \\
 \\
  \\
  \\
  \\
  \\
$\Phi^{-1}(0.75)$ &-0.633&	-0.413&	-0.745&	-0.790&-0.379&	-0.102&	-0.334&	-0.467&-0.504&	-0.336&	-0.462&	-0.525 &-0.438&	-0.327&	-0.454	&-0.479
 \\
 \\
  \\
  \\
  \\
  \\
$\Phi^{-1}(0.9)$ & 0.448&	-0.186&	-0.542&	-0.490&-0.272&	-0.052&	-0.399&	-0.594&-0.519&	-0.324&	-0.510&	-0.642&-0.264&	-0.311&	-0.283&	-0.374
\end{tabular}
\end{center}
\footnotesize{$^a$ $n=3,000$ with cumulative incidence of 0.5. $^b$ The estimates were obtained under unknown nuisance parameters which were estimated by a validation sample of size 500. $^c$ The values in the cells are relative bias of the median of $\beta$, i.e., (median-0.405)/0.405. $^d$ The values in the cells are relative bias of the median of $\omega$, i.e., (median-0.693)/0.693.}
\end{landscape}

\begin{landscape}
{\bf Table 4. Median Relative Bias, Corelation=0.8, Rare Disease$^a$, Validation Sample Size=500$^b$} 
\begin{center}
\fontsize{7.5}{0.3}\selectfont

\begin{tabular}{m{1.09cm}|  m{0.74cm} m{0.74cm} m{0.74cm}  m{0.74cm}| m{0.74cm} m{0.74cm} m{0.74cm} m{0.74cm}| m{0.74cm} m{0.74cm}  m{0.74cm} m{0.74cm}| m{0.74cm} m{0.74cm}  m{0.74cm} m{0.74cm}}
\mc{17}{c}{\textbf{Finite Sample Bias$^c$ in \boldmath{$\beta$}, (\boldmath{$\beta$}, \boldmath{$\omega$})=(0.405, 0.693)}} \\
\\
\\
\\
$\tau$&&&RC1 &&&&RC2 &&&& RR1&&&& RR2\\
\hline
\\
\\
 & EV\_X &  IV\_X & EV\_RM & IV\_RM &  EV\_X &  IV\_X & EV\_RM & IV\_RM&  EV\_X &  IV\_X & EV\_RM & IV\_RM& EV\_X &  IV\_X & EV\_RM & IV\_RM\\
\\
\\
$\Phi^{-1}(0.1)$ &0.617&	0.710&	0.633&	0.626&0.890&	0.853&	0.931&	0.941&0.208&	0.132	&0.233	&0.177&-0.884&	-0.841	&-0.838&	-0.880\\
 \\
 \\
  \\
  \\
  \\
  \\
  $\Phi^{-1}(0.25)$ &0.534&	0.518&	0.513&	0.562& 0.592&	0.576&	0.550&	0.569 &0.109&	0.083&	0.098	&0.106 & -0.040&	-0.022&	-0.004&	-0.013
 \\
 \\
  \\
  \\
  \\
  \\
$\Phi^{-1}(0.5)$ &0.459	&0.468&	0.459&	0.463& 0.324&	0.332&	0.331&	0.326 &0.027	&0.046&	0.031	&0.037&0.008	&0.028	&0.009	&0.008
 \\
 \\
  \\
  \\
  \\
  \\
$\Phi^{-1}(0.75)$ &0.347&	0.351&	0.340&	0.350& 0.157&	0.163&	0.159&	0.164&0.010&	0.017&	0.013	&0.016&0.000 &	0.012	&0.001&0.002\\
 \\
 \\
  \\
  \\
  \\
  \\
$\Phi^{-1}(0.9)$ &0.210&	0.210&	0.205&	0.212 & 0.067&	0.070&	0.062&	0.066&0.003 &	0.007	&0.003 &	0.005&-0.002&	0.007	&-0.004	&0.000
 \\
\\
\\
\\
 \\
 \\
 \\
  \\
  \\
  \\
  \\
  \\

\mc{17}{c}{\textbf{Finite Sample Bias$^d$ in \boldmath{$\omega$}, (\boldmath{$\beta$}, \boldmath{$\omega$})=(0.405, 0.693)}} \\
\\
\\
\\
$\tau$&&&RC1 &&&&RC2 &&&& RR1&&&& RR2\\
\hline
\\
\\
 & EV\_X &  IV\_X & EV\_RM & IV\_RM &  EV\_X &  IV\_X & EV\_RM & IV\_RM&  EV\_X &  IV\_X & EV\_RM & IV\_RM& EV\_X &  IV\_X & EV\_RM & IV\_RM\\
\\
\\
$\Phi^{-1}(0.1)$& -0.384&	-0.478	&-0.414	&-0.409&-0.546&	-0.529&	-0.579&	-0.559& -0.148	&-0.123&	-0.166&	-0.114& 0.472&	0.454	&0.450&	0.482
 \\
 \\
  \\
  \\
  \\
  \\
  $\Phi^{-1}(0.25)$ &-0.357&	-0.345&	-0.361&	-0.362& -0.344&	-0.350&	-0.348	&-0.341 &-0.076&	-0.062	&-0.090&	-0.086& 0.003 &	0.005	&-0.033 &	-0.025
 \\
 \\
  \\
  \\
  \\
  \\
$\Phi^{-1}(0.5)$ &-0.320&	-0.331	&-0.329&	-0.334&-0.170&	-0.187&	-0.181	&-0.188 &-0.042&	-0.060&	-0.057&	-0.063&-0.033	&-0.048&	-0.053&	-0.056
 \\
 \\
  \\
  \\
  \\
  \\
$\Phi^{-1}(0.75)$ & -0.231&	-0.248&	-0.238&	-0.243& 0.002&	-0.026&	-0.016&	-0.027&-0.030&	-0.047&	-0.043	&-0.057&-0.021&	-0.038	&-0.038&	-0.050
 \\
 \\
  \\
  \\
  \\
  \\
$\Phi^{-1}(0.9)$ &-0.083&	-0.110&	-0.096&	-0.091& 0.123&	0.095&	0.116&	0.115&-0.040&	-0.056&	-0.040	&-0.051& -0.035&	-0.048&	-0.038&	-0.045

\end{tabular}
\end{center}
\footnotesize{$^a$ $n=50,000$ with cumulative incidence of 0.03. $^b$ The estimates were obtained under unknown nuisance parameters which were estimated by a validation sample of size 500. $^c$ The values in the cells are relative bias of the median of $\beta$, i.e., (median-0.405)/0.405. $^d$ The values in the cells are relative bias of the median of $\omega$, i.e., (median-0.693)/0.693.}
\end{landscape}
\begin{landscape}
 {\bf Table 5. Median Relative Bias, Corelation=0.6, Rare Disease$^a$, Validation Sample Size=500$^b$} 
\begin{center}
\fontsize{7.5}{0.3}\selectfont

\begin{tabular}{m{1.09cm}|  m{0.74cm} m{0.74cm} m{0.74cm}  m{0.74cm}| m{0.74cm} m{0.74cm} m{0.74cm} m{0.74cm}| m{0.74cm} m{0.74cm}  m{0.74cm} m{0.74cm}| m{0.74cm} m{0.74cm}  m{0.74cm} m{0.74cm}}
\mc{17}{c}{\textbf{Finite Sample Bias$^c$ in \boldmath{$\beta$}, (\boldmath{$\beta$}, \boldmath{$\omega$})=(0.405, 0.693)}} \\
\\
\\
\\
$\tau$&&&RC1 &&&&RC2 &&&& RR1&&&& RR2\\
\hline
\\
\\
 & EV\_X &  IV\_X & EV\_RM & IV\_RM &  EV\_X &  IV\_X & EV\_RM & IV\_RM&  EV\_X &  IV\_X & EV\_RM & IV\_RM& EV\_X &  IV\_X & EV\_RM & IV\_RM\\
\\
$\Phi^{-1}(0.1)$ &0.684&	0.991&	0.590	&0.523 &1.484&	1.597	&1.291&	1.419&-0.460&	-0.612&	-0.488	&-0.399&-2.593&	-2.488	&-2.314&	-2.203\\
 \\
 \\
  \\
  \\
  \\
  \\
  $\Phi^{-1}(0.25)$ &0.826&	0.853&	0.811	&0.849& 0.972&	1.022&	0.887	&0.925& 0.243&	0.184	&0.153&	0.173&-0.334&	-0.345&	-0.370	&-0.352 \\
 \\
 \\
  \\
  \\
  \\
  \\
$\Phi^{-1}(0.5)$ &0.800&	0.830&	0.775&	0.795 &0.551&	0.606	&0.538&	0.539&0.073&	0.094	&0.052&	0.061&0.009&	0.020&	-0.001&	0.002  \\
 \\
 \\
  \\
  \\
  \\
  \\
$\Phi^{-1}(0.75)$ &0.603&	0.619&	0.598&	0.608&0.246&	0.289&	0.244&	0.258&0.018&	0.037&	0.014	&0.030 &-0.004&	0.011	&-0.012&	0.001  \\
 \\
 \\
  \\
  \\
  \\
  \\
$\Phi^{-1}(0.9)$ & 0.351&	0.360	&0.338&	0.352&0.094 &	0.134	&0.086 &	0.102&0.005&	0.023&	-0.001&	0.017&-0.011&	-0.002&	-0.023&	-0.004\\
\\
\\
\\
 \\
 \\
 \\
  \\
  \\
  \\
  \\
  \\

\mc{17}{c}{\textbf{Finite Sample Bias$^d$ in \boldmath{$\omega$}, (\boldmath{$\beta$}, \boldmath{$\omega$})=(0.405, 0.693)}} \\
\\
\\
\\
$\tau$&&&RC1 &&&&RC2 &&&& RR1&&&& RR2\\
\hline
\\
\\
 & EV\_X &  IV\_X & EV\_RM & IV\_RM &  EV\_X &  IV\_X & EV\_RM & IV\_RM&  EV\_X &  IV\_X & EV\_RM & IV\_RM& EV\_X &  IV\_X & EV\_RM & IV\_RM\\
\\
\\
$\Phi^{-1}(0.1)$& -0.489&	-0.631&	-0.455&	-0.455& -0.912&	-0.988&	-0.853&	-0.892& 0.242&	0.282&	0.224	&0.149 & 1.465	&1.432&	1.209&	1.199   \\

 \\
 \\
  \\
  \\
  \\
  \\
  $\Phi^{-1}(0.25)$ &-0.561&	-0.582&	-0.582&	-0.603&-0.601&	-0.635&	-0.560&	-0.584 &-0.188&	-0.167	&-0.164	&-0.180&0.141&	0.145&	0.109	&0.113 \\
 \\
 \\
  \\
  \\
  \\
  \\
$\Phi^{-1}(0.5)$ &-0.574&	-0.597&	-0.556&	-0.567&-0.287&	-0.352	&-0.272&	-0.305  &-0.095&	-0.099&	-0.081	&-0.109 &-0.061&	-0.070&	-0.080	&-0.092\\
 \\
 \\
  \\
  \\
  \\
  \\
$\Phi^{-1}(0.75)$ &-0.413&	-0.459&	-0.413&	-0.421&0.047&	-0.037&	0.054&	0.034&-0.062&	-0.067&	-0.073	&-0.084&-0.060&	-0.058	&-0.078	&-0.090 \\
 \\
 \\
  \\
  \\
  \\
  \\
$\Phi^{-1}(0.9)$ &-0.038&	-0.251	&-0.040&	-0.095& 0.357&	0.198&	0.347	&0.296&-0.054&	-0.063	&-0.054	&-0.090&-0.028&	-0.048&	-0.071	&-0.083 \\

\end{tabular}
\end{center}
\footnotesize{$^a$ $n=50,000$ with cumulative incidence of 0.03. $^b$ The estimates were obtained under unknown nuisance parameters which were estimated by a validation sample of size 500. $^c$ The values in the cells are relative bias of the median of $\beta$, i.e., (median-0.405)/0.405. $^d$ The values in the cells are relative bias of the median of $\omega$, i.e., (median-0.693)/0.693.}
\end{landscape}

\begin{landscape}
{\bf Table 6. Median Relative Bias, Corelation=0.4, Rare Disease$^a$, Validation Sample Size=500$^b$} 
\begin{center}
\fontsize{7.5}{0.3}\selectfont

\begin{tabular}{m{1.09cm}|  m{0.74cm} m{0.74cm} m{0.74cm}  m{0.74cm}| m{0.74cm} m{0.74cm} m{0.74cm} m{0.74cm}| m{0.74cm} m{0.74cm}  m{0.74cm} m{0.74cm}| m{0.74cm} m{0.74cm}  m{0.74cm} m{0.74cm}}
\mc{17}{c}{\textbf{Finite Sample Bias$^c$ in \boldmath{$\beta$}, (\boldmath{$\beta$}, \boldmath{$\omega$})=(0.405, 0.693)}} \\
\\
\\
\\
$\tau$&&&RC1 &&&&RC2 &&&& RR1&&&& RR2\\
\hline
\\
\\
 & EV\_X &  IV\_X & EV\_RM & IV\_RM &  EV\_X &  IV\_X & EV\_RM & IV\_RM&  EV\_X &  IV\_X & EV\_RM & IV\_RM& EV\_X &  IV\_X & EV\_RM & IV\_RM\\
\\
\\
$\Phi^{-1}(0.1)$ &-3.909&	-0.196&	0.022&	0.115&1.663	&2.220&	1.240&	1.338&-2.638&	-2.456&	-1.528&	-1.143&-5.216&	-4.979&	-3.585	&-2.967
 \\
 \\
  \\
  \\
  \\
  \\
  $\Phi^{-1}(0.25)$ &1.072&	1.216&	0.830	&0.791 & 1.167&	1.653&	1.016&	0.969 & -0.342&	-0.369	&-0.259&	-0.363&-1.933&	-1.600&	-1.368	&-1.419
 \\
 \\
  \\
  \\
  \\
  \\
$\Phi^{-1}(0.5)$ &1.035&	1.134	&0.980&	1.044& 0.664&	0.962&	0.603&	0.622 &0.073&	0.134	&0.004	&0.073&-0.313&	-0.348&	-0.385	&-0.345
 \\
 \\
  \\
  \\
  \\
  \\
$\Phi^{-1}(0.75)$ &0.775	&0.819&	0.727	&0.739 & 0.276&	0.485&	0.252&	0.288&0.026	&0.077	&-0.005	&0.027 &-0.058	&-0.125	&-0.192	&-0.143
 \\
 \\
  \\
  \\
  \\
  \\
$\Phi^{-1}(0.9)$ & 0.377&	0.410	&0.207	&0.193& 0.098&	0.238&	0.069&	0.094&-0.009&	0.046&	-0.014&	0.042& -0.048&	-0.061	&-0.199	&-0.137
\\
\\
\\
\\
 \\
 \\
 \\
  \\
  \\
  \\
  \\
  \\

\mc{17}{c}{\textbf{Finite Sample Bias$^d$ in \boldmath{$\omega$}, (\boldmath{$\beta$}, \boldmath{$\omega$})=(0.405, 0.693)}} \\
\\
\\
\\
$\tau$&&&RC1 &&&&RC2 &&&& RR1&&&& RR2\\
\hline
\\
\\
 & EV\_X &  IV\_X & EV\_RM & IV\_RM &  EV\_X &  IV\_X & EV\_RM & IV\_RM&  EV\_X &  IV\_X & EV\_RM & IV\_RM& EV\_X &  IV\_X & EV\_RM & IV\_RM\\
\\
\\
$\Phi^{-1}(0.1)$& 1.998&	-0.119&	-0.452&	-0.577 & -1.046&	-1.524&	-0.893&	-0.972& 1.487&	1.338&	0.700	&0.577& 3.027	&2.805&	1.738&	1.556
 \\
 \\
  \\
  \\
  \\
  \\
  $\Phi^{-1}(0.25)$ &-0.718	&-0.872&	-0.779&	-0.749& -0.730&	-1.122&	-0.725&	-0.722&0.140&	0.143	&-0.013	&0.015&1.046&	0.786	&0.501	&0.561
 \\
 \\
  \\
  \\
  \\
  \\
$\Phi^{-1}(0.5)$ &-0.750&	-0.899&	-0.742&	-0.747& -0.315&	-0.740&	-0.349&	-0.395&-0.116&	-0.123&	-0.131&	-0.176&0.119	&0.126&	-0.052	&-0.078
 \\
 \\
  \\
  \\
  \\
  \\
$\Phi^{-1}(0.75)$ &-0.484	&-0.896&	-0.541&	-0.568 & 0.130&	-0.422	&0.107&	0.079 &-0.104&	-0.092&	-0.090&	-0.126&-0.020&	0.049	&-0.154&	-0.176
 \\
 \\
  \\
  \\
  \\
  \\
$\Phi^{-1}(0.9)$ & 0.295&	-0.961	&-0.181&	-0.331& 0.634&	-0.237	&0.444&	0.231 &-0.078	&-0.123&	-0.055&	-0.182 &0.069&	0.097&	-0.074&	-0.088

\end{tabular}
\end{center}
\footnotesize{$^a$ $n=50,000$ with cumulative incidence of 0.03. $^b$ The estimates were obtained under unknown nuisance parameters which were estimated by a validation sample of size 500. $^c$ The values in the cells are relative bias of the median of $\beta$, i.e., (median-0.405)/0.405. $^d$ The values in the cells are relative bias of the median of $\omega$, i.e., (median-0.693)/0.693.}
\end{landscape}

\begin{landscape}
\subsection*{B.2. \enspace Mean Square Error (MSE)}
{\bf Table 1. MSE, Corelation=0.8, Common Disease$^a$, Validation Sample Size=500$^b$}
\begin{center}
\fontsize{7.5}{0.3}\selectfont
\begin{tabular}{m{1.09cm}|  m{0.74cm} m{0.74cm} m{0.74cm}  m{0.74cm}| m{0.74cm} m{0.74cm} m{0.74cm} m{0.74cm}| m{0.74cm} m{0.74cm}  m{0.74cm} m{0.74cm}| m{0.74cm} m{0.74cm}  m{0.74cm} m{0.74cm}}
\mc{17}{c}{\textbf{Finite Sample MSE$^c$ in \boldmath{$\beta$}, (\boldmath{$\beta$}, \boldmath{$\omega$})=(0.405, 0.693)}} \\
\\
\\
\\
$\tau$&&&RC1 &&&&RC2 &&&& RR1&&&& RR2\\
\hline
\\
\\
 & EV\_X &  IV\_X & EV\_RM & IV\_RM &  EV\_X &  IV\_X & EV\_RM & IV\_RM&  EV\_X &  IV\_X & EV\_RM & IV\_RM& EV\_X &  IV\_X & EV\_RM & IV\_RM\\
\\
\\
$\Phi^{-1}(0.1)$ &0.462&	0.322&	0.432	&0.489&0.393&	0.253&	0.359&	0.433&0.751	&0.493	&0.616&	0.905&0.667&	0.415&	0.529	&0.851\\
 \\
 \\
  \\
  \\
  \\
  \\
  $\Phi^{-1}(0.25)$ & 0.067&	0.054&	0.059	&0.071&0.067&	0.044&	0.063	&0.076&0.071	&0.046&	0.060&	0.067 &0.066&	0.044&	0.056	&0.061 \\
 \\
 \\
  \\
  \\
  \\
  \\
$\Phi^{-1}(0.5)$ & 0.019	&0.015	&0.019&	0.024 &0.012	&0.008&	0.012&	0.015&0.013&	0.009	&0.012	&0.013&0.013	&0.009	&0.012	&0.013  \\
 \\
 \\
  \\
  \\
  \\
  \\
$\Phi^{-1}(0.75)$ &0.007&	0.005	&0.007&	0.010&0.004&	0.003&	0.004&	0.005&0.005&	0.004	&0.005&	0.005&0.005	&0.004	&0.005&	0.005
 \\
 \\
  \\
  \\
  \\
  \\
$\Phi^{-1}(0.9)$ &0.003&	0.002&	0.003&	0.004 &0.002&	0.002	&0.002&	0.003&0.003&	0.002	&0.003&	0.003&0.003&	0.002	&0.003	&0.003 \\
\\
\\
\\
 \\
 \\
 \\
  \\
  \\
  \\
  \\
  \\

\mc{17}{c}{\textbf{Finite Sample MSE$^d$ in \boldmath{$\omega$}, (\boldmath{$\beta$}, \boldmath{$\omega$})=(0.405, 0.693)}} \\
\\
\\
\\
$\tau$&&&RC1 &&&&RC2 &&&& RR1&&&& RR2\\
\hline
\\
\\
 & EV\_X &  IV\_X & EV\_RM & IV\_RM &  EV\_X &  IV\_X & EV\_RM & IV\_RM&  EV\_X &  IV\_X & EV\_RM & IV\_RM& EV\_X &  IV\_X & EV\_RM & IV\_RM\\
\\
\\
$\Phi^{-1}(0.1)$& 0.577&	0.411&	0.541&	0.594& 0.556&	0.342&	0.521	&0.579& 0.868	&0.586	&0.718	&1.016& 0.712&	0.467&	0.566	&0.889    \\

 \\
 \\
  \\
  \\
  \\
  \\
  $\Phi^{-1}(0.25)$ &0.169	&0.122 &	0.156 &	0.154& 0.156&	0.093&	0.148&	0.150&0.128&	0.090&	0.113	&0.114&0.115	&0.084&	0.103&	0.104
 \\
 \\
  \\
  \\
  \\
  \\
$\Phi^{-1}(0.5)$ &0.114	&0.077&	0.109&	0.109&0.062&	0.038&	0.059&	0.059&0.052&	0.042	&0.049	&0.048&0.052&	0.041&	0.050	&0.048\\
 \\
 \\
  \\
  \\
  \\
  \\

$\Phi^{-1}(0.75)$ &0.090&	0.058	&0.093	&0.083&0.047&	0.029	&0.048&	0.044 &0.058	&0.042	&0.057	&0.052 &0.058	&0.044	&0.058	&0.053 \\
 \\
 \\
  \\
  \\
  \\
  \\
$\Phi^{-1}(0.9)$ &0.094&	0.060&	0.109&	0.100&0.080&	0.053&	0.091&	0.083&0.088&	0.066&	0.098&	0.086& 0.090	&0.066&	0.099&	0.088  \\
\end{tabular}
\end{center}
\footnotesize{$^a$ $n=3,000$ with cumulative incidence of 0.5. $^b$ The estimates were obtained under unknown nuisance parameters which were estimated by a validation sample of size 500. $^c$ The values in the cells are the MSE of $\beta$. $^d$ The values in the cells are the MSE of $\omega$.}
\end{landscape}

\begin{landscape}
{\bf Table 2. MSE, Corelation=0.6, Common Disease$^a$, Validation Sample Size=500$^b$} 
\begin{center}
\fontsize{7.5}{0.3}\selectfont

\begin{tabular}{m{1.09cm}|  m{0.74cm} m{0.74cm} m{0.74cm}  m{0.74cm}| m{0.74cm} m{0.74cm} m{0.74cm} m{0.74cm}| m{0.74cm} m{0.74cm}  m{0.74cm} m{0.74cm}| m{0.74cm} m{0.74cm}  m{0.74cm} m{0.74cm}}
\mc{17}{c}{\textbf{Finite Sample MSE$^c$ in \boldmath{$\beta$}, (\boldmath{$\beta$}, \boldmath{$\omega$})=(0.405, 0.693)}} \\
\\
\\
\\
$\tau$&&&RC1 &&&&RC2 &&&& RR1&&&& RR2\\
\hline
\\
\\
 & EV\_X &  IV\_X & EV\_RM & IV\_RM &  EV\_X &  IV\_X & EV\_RM & IV\_RM&  EV\_X &  IV\_X & EV\_RM & IV\_RM& EV\_X &  IV\_X & EV\_RM & IV\_RM\\
\\
$\Phi^{-1}(0.1)$ &2.111	&0.871	&2.257&	2.179&0.911&	0.312	&0.833&	1.088 & 1.758&	0.979	&1.383&	1.680&2.895&	0.844	&2.229&	2.890\\
 \\
 \\
  \\
  \\
  \\
  \\
  $\Phi^{-1}(0.25)$ &0.160&	0.108&	0.152	&0.1720 &0.163&	0.056	&0.151&	0.208 &0.404&	0.102	&0.385	&0.481&0.329&	0.091	&0.338&	0.427\\
 \\
 \\
  \\
  \\
  \\
  \\
$\Phi^{-1}(0.5)$ &0.039&	0.028&	0.041&	0.057  &0.030&	0.013&	0.032&	0.046&0.042	&0.016	&0.042&	0.050 &0.042&	0.016	&0.037	&0.044\\
 \\
 \\
  \\
  \\
  \\
  \\
$\Phi^{-1}(0.75)$ &0.014&	0.010	&0.017	&0.025&0.010&	0.005	&0.012&	0.015 &0.014	&0.006&	0.016&	0.016&0.014&	0.006&	0.015&	0.016 \\
 \\
 \\
  \\
  \\
  \\
  \\
$\Phi^{-1}(0.9)$ &0.005&	0.003&	0.007	&0.009&0.006&	0.003&	0.008	&0.009&0.007&	0.004&	0.009	&0.010 &0.007	&0.004	&0.009	&0.009  \\
\\
\\
\\
 \\
 \\
 \\
  \\
  \\
  \\
  \\
  \\

\mc{17}{c}{\textbf{Finite Sample MSE$^d$ in \boldmath{$\omega$}, (\boldmath{$\beta$}, \boldmath{$\omega$})=(0.405, 0.693)}} \\
\\
\\
\\
$\tau$&&&RC1 &&&&RC2 &&&& RR1&&&& RR2\\
\hline
\\
\\
 & EV\_X &  IV\_X & EV\_RM & IV\_RM &  EV\_X &  IV\_X & EV\_RM & IV\_RM&  EV\_X &  IV\_X & EV\_RM & IV\_RM& EV\_X &  IV\_X & EV\_RM & IV\_RM\\
\\
\\
$\Phi^{-1}(0.1)$& 2.359	&1.056&	2.549&	2.390& 1.321&	0.399&	1.194&	1.421& 1.966	&1.163	&1.598	&1.862& 2.826	&0.950	&2.175	&2.811   \\

 \\
 \\
  \\
  \\
  \\
  \\
  $\Phi^{-1}(0.25)$ &0.372&	0.249&	0.353	&0.348&0.393&	0.117&	0.360&	0.423&0.588	&0.208	&0.547&	0.653&0.444&	0.188	&0.442	&0.523\\

 \\
 \\
  \\
  \\
  \\
  \\
$\Phi^{-1}(0.5)$ &0.259	&0.152	&0.248&	0.250 &0.168	&0.050&	0.162&	0.183 &0.156&	0.089	&0.143	&0.155&0.147&	0.089&	0.138	&0.146\\
 \\
 \\
  \\
  \\
  \\
  \\
$\Phi^{-1}(0.75)$ &0.203&	0.101&	0.225&	0.231&0.147&	0.040&	0.155&	0.165 &0.147	&0.083	&0.140	&0.156&0.147&	0.084&	0.141	&0.150\\
 \\
 \\
  \\
  \\
  \\
  \\

$\Phi^{-1}(0.9)$ &0.603&	0.090&	0.911&	0.933 &0.294&	0.079&	0.388&	0.380&0.227	&0.108&	0.254&	0.267 &0.209&	0.107	&0.225&	0.236\\

\end{tabular}
\end{center}
\footnotesize{$^a$ $n=3,000$ with cumulative incidence of 0.5. $^b$ The estimates were obtained under unknown nuisance parameters which were estimated by a validation sample of size 500. $^c$ The values in the cells are the MSE of $\beta$. $^d$ The values in the cells are the MSE of $\omega$.}
\end{landscape}

\begin{landscape}
{\bf Table 3. MSE, Corelation=0.4, Common Disease$^a$, Validation Sample Size=500$^b$}
\begin{center}
\fontsize{7.5}{0.3}\selectfont
\begin{tabular}{m{1.09cm}|  m{0.74cm} m{0.74cm} m{0.74cm}  m{0.74cm}| m{0.74cm} m{0.74cm} m{0.74cm} m{0.74cm}| m{0.74cm} m{0.74cm}  m{0.74cm} m{0.74cm}| m{0.74cm} m{0.74cm}  m{0.74cm} m{0.74cm}}
\mc{17}{c}{\textbf{Finite Sample MSE$^c$ in \boldmath{$\beta$}, (\boldmath{$\beta$}, \boldmath{$\omega$})=(0.405, 0.693)}} \\
\\
\\
\\
$\tau$&&&RC1 &&&&RC2 &&&& RR1&&&& RR2\\
\hline
\\
\\
 & EV\_X &  IV\_X & EV\_RM & IV\_RM &  EV\_X &  IV\_X & EV\_RM & IV\_RM&  EV\_X &  IV\_X & EV\_RM & IV\_RM& EV\_X &  IV\_X & EV\_RM & IV\_RM\\
\\
\\
$\Phi^{-1}(0.1)$ &5.767&	1.310&	4.033&	3.544&2.338	&0.563	&2.270&	2.434&2.903 &	1.149 &	2.672 &	2.968 &6.807&	1.020	&6.075&	7.117\\
 \\
 \\
  \\
  \\
  \\
  \\
  $\Phi^{-1}(0.25)$ &0.853&	0.195&	1.201&	1.365 &0.507&	0.094	&0.751&	0.852&1.318 &	0.153	&1.277	&1.439 &1.949	&0.138&	2.069	&2.410\\
 \\
 \\
  \\
  \\
  \\
  \\
$\Phi^{-1}(0.5)$ &0.061&	0.046&	0.134&	0.167&0.118&	0.024&	0.229&	0.248 &0.343 &	0.024	&0.403	&0.523&0.315	&0.024	&0.442	&0.572\\
 \\
 \\
  \\
  \\
  \\
  \\
$\Phi^{-1}(0.75)$ &0.021&	0.014&	0.036	&0.049&0.042&	0.009&	0.078	&0.086 &0.069	&0.009	&0.113	&0.135&0.065&	0.009	&0.103	&0.136 \\
 \\
 \\
  \\
  \\
  \\
  \\
$\Phi^{-1}(0.9)$ &0.005	&0.005&	0.008&	0.007&0.023&	0.006&	0.039&	0.044 &0.030 &	0.006 &	0.051 &	0.053&0.033	&0.007&	0.060	&0.062  \\
\\
\\
\\
 \\
 \\
 \\
  \\
  \\
  \\
  \\
  \\

\mc{17}{c}{\textbf{Finite Sample MSE$^d$ in \boldmath{$\omega$}, (\boldmath{$\beta$}, \boldmath{$\omega$})=(0.405, 0.693)}} \\
\\
\\
\\
$\tau$&&&RC1 &&&&RC2 &&&& RR1&&&& RR2\\
\hline
\\
\\
 & EV\_X &  IV\_X & EV\_RM & IV\_RM &  EV\_X &  IV\_X & EV\_RM & IV\_RM&  EV\_X &  IV\_X & EV\_RM & IV\_RM& EV\_X &  IV\_X & EV\_RM & IV\_RM\\
\\
\\
$\Phi^{-1}(0.1)$& 5.454&	1.577&	4.064&	3.848& 3.190	&0.558&	3.023	&3.279& 2.809 &	1.400	&2.847 &	3.093& 6.275&	1.191&	5.556&	6.595 \\
 \\
 \\
  \\
  \\
  \\
  \\
  $\Phi^{-1}(0.25)$ &1.168&	0.387	&1.512	&1.664&1.066&	0.122	&1.345	&1.468&1.696	&0.294	&1.712 &	1.964&2.032&	0.281&	2.234	&2.736\\
 \\
 \\
  \\
  \\
  \\
  \\
$\Phi^{-1}(0.5)$ &0.392&	0.210&	0.394	&0.404 &0.524	&0.054 &	0.787 &	0.799&0.708 &	0.133 &	0.801 &	0.979 &0.595&	0.142	&0.797&	0.960\\
 \\
 \\
  \\
  \\
  \\
  \\
$\Phi^{-1}(0.75)$ &0.569&	0.116&	1.270&	1.304&0.620&	0.049&	0.974&	1.144&0.475	&0.105	&0.623	&0.732 &0.401	&0.114&	0.528	&0.705
 \\
 \\
  \\
  \\
  \\
  \\
$\Phi^{-1}(0.9)$ & 6.865&	0.102&	3.239&	3.764&1.667&	0.096	&2.167	&2.681 &0.990	&0.143 &	1.105 &	1.320& 0.908&	0.158	&1.015	&1.321 \\
\end{tabular}
\end{center}
\footnotesize{$^a$ $n=3,000$ with cumulative incidence of 0.5. $^b$ The estimates were obtained under unknown nuisance parameters which were estimated by a validation sample of size 500. $^c$ The values in the cells are the MSE of $\beta$. $^d$ The values in the cells are the MSE of $\omega$.}
\end{landscape}

\begin{landscape}
{\bf Table 4. MSE, Corelation=0.8, Rare Disease$^a$, Validation Sample Size=500$^b$} 
\begin{center}
\fontsize{7.5}{0.3}\selectfont

\begin{tabular}{m{1.09cm}|  m{0.74cm} m{0.74cm} m{0.74cm}  m{0.74cm}| m{0.74cm} m{0.74cm} m{0.74cm} m{0.74cm}| m{0.74cm} m{0.74cm}  m{0.74cm} m{0.74cm}| m{0.74cm} m{0.74cm}  m{0.74cm} m{0.74cm}}
\mc{17}{c}{\textbf{Finite Sample MSE$^c$ in \boldmath{$\beta$}, (\boldmath{$\beta$}, \boldmath{$\omega$})=(0.405, 0.693)}} \\
\\
\\
\\
$\tau$&&&RC1 &&&&RC2 &&&& RR1&&&& RR2\\
\hline
\\
\\
 & EV\_X &  IV\_X & EV\_RM & IV\_RM &  EV\_X &  IV\_X & EV\_RM & IV\_RM&  EV\_X &  IV\_X & EV\_RM & IV\_RM& EV\_X &  IV\_X & EV\_RM & IV\_RM\\
\\
\\
$\Phi^{-1}(0.1)$ &0.833	&0.861&	0.737&	0.894&0.569	&0.521&	0.540&	0.636&1.113&	0.915	&1.060	&1.064 &1.330	&1.043&	1.283	&1.259\\
 \\
 \\
  \\
  \\
  \\
  \\
  $\Phi^{-1}(0.25)$ &0.114&	0.107	&0.105&	0.124&0.120&	0.118&	0.109&	0.122 &0.112&	0.109&	0.104&	0.131&0.098&	0.095	&0.087&	0.110 \\
 \\
 \\
  \\
  \\
  \\
  \\
$\Phi^{-1}(0.5)$ &0.045&	0.047&	0.044	&0.046&0.031	&0.031&	0.030&	0.031 &0.017	&0.017&	0.015	&0.017&0.017&	0.017	&0.014	&0.017 \\
 \\
 \\
  \\
  \\
  \\
  \\
$\Phi^{-1}(0.75)$ &0.023&	0.024&	0.023	&0.024 &0.009 &	0.009 &	0.008	&0.009&0.005	&0.005&	0.005&	0.005&0.005&	0.005&	0.005&	0.005 \\
 \\
 \\
  \\
  \\
  \\
  \\
$\Phi^{-1}(0.9)$ &0.009	&0.009&	0.009&	0.010&0.003&	0.003&	0.003&	0.003&0.003	&0.003&	0.003	&0.003&0.003&	0.003	&0.003	&0.003\\
\\
\\
\\
 \\
 \\
 \\
  \\
  \\
  \\
  \\
  \\

\mc{17}{c}{\textbf{Finite Sample MSE$^d$ in \boldmath{$\omega$}, (\boldmath{$\beta$}, \boldmath{$\omega$})=(0.405, 0.693)}} \\
\\
\\
\\
$\tau$&&&RC1 &&&&RC2 &&&& RR1&&&& RR2\\
\hline
\\
\\
 & EV\_X &  IV\_X & EV\_RM & IV\_RM &  EV\_X &  IV\_X & EV\_RM & IV\_RM&  EV\_X &  IV\_X & EV\_RM & IV\_RM& EV\_X &  IV\_X & EV\_RM & IV\_RM\\
\\
\\
$\Phi^{-1}(0.1)$& 0.847&	0.899	&0.762&	0.920& 0.598	&0.556&	0.579&	0.664& 1.139&	0.943	&1.089&	1.094 & 1.337	&1.053	&1.293	&1.276   \\

 \\
 \\
  \\
  \\
  \\
  \\
  $\Phi^{-1}(0.25)$ &0.135&	0.128&	0.132&	0.143&0.130&	0.136&	0.131	&0.140&0.125	&0.126	&0.121	&0.150&0.110&	0.112	&0.102	&0.127 \\

 \\
 \\
  \\
  \\
  \\
  \\
$\Phi^{-1}(0.5)$ &0.065&	0.069&	0.067&	0.070&0.038	&0.041&	0.038&	0.042&0.028	&0.030	&0.027	&0.030&0.027&	0.030	&0.026	&0.030\\
 \\
 \\
  \\
  \\
  \\
  \\
$\Phi^{-1}(0.75)$ &0.037 &	0.042	&0.038	&0.040&0.018&	0.020&	0.017	&0.020& 0.016	&0.018	&0.016	&0.018& 0.016&	0.018	&0.016	&0.018 \\
 \\
 \\
  \\
  \\
  \\
  \\
$\Phi^{-1}(0.9)$ &0.026&	0.031	&0.029&	0.030&0.039	&0.039&	0.039&	0.042&0.022&	0.024&	0.022&	0.024&0.022	&0.023	&0.022&	0.024 \\

\end{tabular}
\end{center}
\footnotesize{$^a$ $n=50,000$ with cumulative incidence of 0.03. $^b$ The estimates were obtained under unknown nuisance parameters which were estimated by a validation sample of size 500. $^c$ The values in the cells are the MSE of $\beta$. $^d$ The values in the cells are the MSE of $\omega$.}
\end{landscape}
\begin{landscape}
{\bf Table 5. MSE, Corelation=0.6, Rare Disease$^a$, Validation Sample Size=500$^b$}
\begin{center}
\fontsize{7.5}{0.3}\selectfont
\begin{tabular}{m{1.09cm}|  m{0.74cm} m{0.74cm} m{0.74cm}  m{0.74cm}| m{0.74cm} m{0.74cm} m{0.74cm} m{0.74cm}| m{0.74cm} m{0.74cm}  m{0.74cm} m{0.74cm}| m{0.74cm} m{0.74cm}  m{0.74cm} m{0.74cm}}
\mc{17}{c}{\textbf{Finite Sample MSE$^c$ in \boldmath{$\beta$}, (\boldmath{$\beta$}, \boldmath{$\omega$})=(0.405, 0.693)}} \\
\\
\\
\\
$\tau$&&&RC1 &&&&RC2 &&&& RR1&&&& RR2\\
\hline
\\
\\
 & EV\_X &  IV\_X & EV\_RM & IV\_RM &  EV\_X &  IV\_X & EV\_RM & IV\_RM&  EV\_X &  IV\_X & EV\_RM & IV\_RM& EV\_X &  IV\_X & EV\_RM & IV\_RM\\
\\
\\
$\Phi^{-1}(0.1)$ &2.655&	2.558&	2.604&	2.665&1.109&	1.191&	1.030	&1.227&1.932	&1.951&	1.788&	2.058&4.255	&4.083	&3.531	&4.124\\
 \\
 \\
  \\
  \\
  \\
  \\
  $\Phi^{-1}(0.25)$ &0.294	&0.299&	0.301&	0.356&0.284&	0.305&	0.258&	0.297&0.634&	0.556&	0.609	&0.687&0.600	&0.514	&0.599	&0.717  \\
 \\
 \\
  \\
  \\
  \\
  \\
$\Phi^{-1}(0.5)$ &0.123&	0.132&	0.118	&0.128&0.082&	0.092	&0.077&	0.084&0.056&	0.057&	0.051&	0.062&0.051	&0.050	&0.043	&0.051\\
 \\
 \\
  \\
  \\
  \\
  \\
$\Phi^{-1}(0.75)$ &0.066&	0.070&	0.069&	0.073&0.021&	0.025&	0.021	&0.026&0.014	&0.015	&0.014	&0.017 &0.014	&0.014&	0.013	&0.016\\
 \\
 \\
  \\
  \\
  \\
  \\
$\Phi^{-1}(0.9)$ &0.024	&0.026	&0.025&	0.028&0.007&	0.010&	0.008&	0.010&0.007&	0.007&	0.008&	0.009 &0.007	&0.007&	0.008	&0.009\\
\\
\\
\\
 \\
 \\
 \\
  \\
  \\
  \\
  \\
  \\

\mc{17}{c}{\textbf{Finite Sample MSE$^d$ in \boldmath{$\omega$}, (\boldmath{$\beta$}, \boldmath{$\omega$})=(0.405, 0.693)}} \\
\\
\\
\\
$\tau$&&&RC1 &&&&RC2 &&&& RR1&&&& RR2\\
\hline
\\
\\
 & EV\_X &  IV\_X & EV\_RM & IV\_RM &  EV\_X &  IV\_X & EV\_RM & IV\_RM&  EV\_X &  IV\_X & EV\_RM & IV\_RM& EV\_X &  IV\_X & EV\_RM & IV\_RM\\
\\
\\
$\Phi^{-1}(0.1)$& 2.735	&2.609&	2.708&	2.813& 1.207&	1.291&	1.176&	1.345& 1.983&	1.990	&1.901&	2.170& 4.258	&4.130	&3.499&	4.166  \\
 \\
 \\
  \\
  \\
  \\
  \\
  $\Phi^{-1}(0.25)$ &0.345&	0.353	&0.368&	0.428&0.343&	0.363	&0.324&	0.369&0.698&	0.620	&0.683&	0.777 &0.652	&0.563	&0.650&	0.788\\
 \\
 \\
  \\
  \\
  \\
  \\
$\Phi^{-1}(0.5)$ &0.184&	0.197	&0.174&	0.182&0.110&	0.124&	0.107	&0.124 &0.092&	0.094&	0.097&	0.111&0.085	&0.086	&0.087	&0.097\\
 \\
 \\
  \\
  \\
  \\
  \\
$\Phi^{-1}(0.75)$ &0.109	&0.127	&0.113&	0.122&0.070&	0.060&	0.082&	0.096 &0.047	&0.050	&0.054	&0.062 &0.047	&0.049	&0.053	&0.060\\
 \\
 \\
  \\
  \\
  \\
  \\
$\Phi^{-1}(0.9)$ &0.178&	0.176&	0.212	&0.316 &0.225&	0.147&	0.279&	0.305&0.063&	0.068	&0.073&	0.086&0.062	&0.066&	0.069	&0.079\\
\end{tabular}
\end{center}
\footnotesize{$^a$ $n=50,000$ with cumulative incidence of 0.03. $^b$ The estimates were obtained under unknown nuisance parameters which were estimated by a validation sample of size 500. $^c$ The values in the cells are the MSE of $\beta$. $^d$ The values in the cells are the MSE of $\omega$.}
\end{landscape}

\begin{landscape}
{\bf Table 6. MSE, Corelation=0.4, Rare Disease$^a$, Validation Sample Size=500$^b$} 
\begin{center}
\fontsize{7.5}{0.3}\selectfont

\begin{tabular}{m{1.09cm}|  m{0.74cm} m{0.74cm} m{0.74cm}  m{0.74cm}| m{0.74cm} m{0.74cm} m{0.74cm} m{0.74cm}| m{0.74cm} m{0.74cm}  m{0.74cm} m{0.74cm}| m{0.74cm} m{0.74cm}  m{0.74cm} m{0.74cm}}
\mc{17}{c}{\textbf{Finite Sample MSE$^a$ in \boldmath{$\beta$}, (\boldmath{$\beta$}, \boldmath{$\omega$})=(0.405, 0.693)}} \\
\\
\\
\\
$\tau$&&&RC1 &&&&RC2 &&&& RR1&&&& RR2\\
\hline
\\
\\
 & EV\_X &  IV\_X & EV\_RM & IV\_RM &  EV\_X &  IV\_X & EV\_RM & IV\_RM&  EV\_X &  IV\_X & EV\_RM & IV\_RM& EV\_X &  IV\_X & EV\_RM & IV\_RM\\
\\
\\
$\Phi^{-1}(0.1)$ &7.118&	2.972	&4.357&	3.920&2.530&	2.637	&2.380	&2.346 & 4.432&	3.604&	4.209	&4.257&11.364	&10.114&	10.754&	11.166 \\
 \\
 \\
  \\
  \\
  \\
  \\
  $\Phi^{-1}(0.25)$ &1.325 &	1.269 &	1.508	&1.542 &0.681&	1.128&	0.853&	0.962&1.579&	1.148	&2.004&	1.765 &2.953&	1.935	&3.949&	3.551   \\
 \\
 \\
  \\
  \\
  \\
  \\
$\Phi^{-1}(0.5)$ & 0.207&	0.286	&0.274&	0.337&0.191	&0.350&	0.245	&0.315& 0.426&	0.373&	0.482	&0.555&0.353	&0.366	&0.522	&0.632  \\
 \\
 \\
  \\
  \\
  \\
  \\
$\Phi^{-1}(0.75)$ &0.109&	0.141	&0.135&	0.140 &0.054&	0.103&	0.082	&0.109&0.066	&0.097&	0.093&	0.132& 0.059	&0.107	&0.072	&0.118 \\
 \\
 \\
  \\
  \\
  \\
  \\
$\Phi^{-1}(0.9)$ &0.029	&0.043	&0.017 &	0.016&0.024	&0.037&	0.040&	0.051& 0.029	&0.036	&0.049	0.065 &0.028	&0.038	&0.045	&0.072   \\
\\
\\
\\
 \\
 \\
 \\
  \\
  \\
  \\
  \\
  \\

\mc{17}{c}{\textbf{Finite Sample MSE$^a$ in \boldmath{$\omega$}, (\boldmath{$\beta$}, \boldmath{$\omega$})=(0.405, 0.693)}} \\
\\
\\
\\
$\tau$&&&RC1 &&&&RC2 &&&& RR1&&&& RR2\\
\hline
\\
\\
 & EV\_X &  IV\_X & EV\_RM & IV\_RM &  EV\_X &  IV\_X & EV\_RM & IV\_RM&  EV\_X &  IV\_X & EV\_RM & IV\_RM& EV\_X &  IV\_X & EV\_RM & IV\_RM\\
\\
\\
$\Phi^{-1}(0.1)$& 6.589	&3.111&	4.597	&4.166& 2.914&	3.049&	2.837&	2.828& 4.509&	3.681	&4.157	&4.210 & 11.510	&10.156	&10.372	&10.757   \\

 \\
 \\
  \\
  \\
  \\
  \\
  $\Phi^{-1}(0.25)$ &1.415 &	1.392	&1.734 &	1.780&0.946&	1.378&	1.265&	1.372&1.757&	1.304	&2.477	&2.206 &3.102	&2.049	&4.269	&3.759  \\

 \\
 \\
  \\
  \\
  \\
  \\
$\Phi^{-1}(0.5)$ &0.321&	0.459&	0.341&	0.357&0.396&	0.599&	0.655&	0.723&0.620&	0.519&	0.831&	0.909& 0.519	&0.536&	0.807&	1.004 \\
 \\
 \\
  \\
  \\
  \\
  \\
$\Phi^{-1}(0.75)$ &0.312	&0.598&	0.715&	0.944&0.432&	0.361	&0.801&	0.916&0.225&	0.274&	0.390	&0.456&0.201	&0.329&	0.327	&0.415 \\
 \\
 \\
  \\
  \\
  \\
  \\
$\Phi^{-1}(0.9)$ & 5.219&	1.374&	3.371&	3.127&1.423	&0.546&	1.828&	1.867&0.339&	0.371&	0.666&	0.754 & 0.255	&0.404&	0.674&	0.833  \\
\\
\\
\\
\\
 \\

\end{tabular}
\end{center}
\footnotesize{$^a$ $n=50,000$ with cumulative incidence of 0.03. $^b$ The estimates were obtained under unknown nuisance parameters which were estimated by a validation sample of size 500. $^c$ The values in the cells are the MSE of $\beta$. $^d$ The values in the cells are the MSE of $\omega$.}
\end{landscape}

\subsection*{B.3. \enspace Convergence Percent (Pctgud)}
{\bf Table 1. Pctgud$^a$, Common Disease$^b$, Validation Sample Size=500$^c$}
\begin{center}
\fontsize{7.5}{0.3}\selectfont
\begin{tabular}{m{1.09cm}|  m{0.74cm} m{0.74cm} m{0.74cm}  m{0.74cm}| m{0.74cm} m{0.74cm} m{0.74cm} m{0.74cm}| m{0.74cm} m{0.74cm}  m{0.74cm} m{0.74cm}}
\mc{13}{c}{\textbf{Corelation=0.8}} \\
\\
\\
\\
$\tau$&&&RC1 &&&&RC2 &&&& RR\\
\hline
\\
\\
 & EV\_X &  IV\_X & EV\_RM & IV\_RM &  EV\_X &  IV\_X & EV\_RM & IV\_RM&  EV\_X &  IV\_X & EV\_RM & IV\_RM\\
\\
\\
$\Phi^{-1}(0.1)$&1&	0.992&	1&	1&1&	1&	1&	1&0.973&	0.997&	0.973&	0.985

 \\
 \\
  \\
  \\
  \\
  \\
  $\Phi^{-1}(0.25)$&1&	1&	1&	1&1&	1&	1&	1&1&	1&	1&	1    \\
 \\
 \\
  \\
  \\
  \\
  \\
$\Phi^{-1}(0.5)$&1&	1&	1&	1&1&	1&	1&	1 &1&	1&	1&	1 \\
 \\
 \\
  \\
  \\
  \\
  \\
$\Phi^{-1}(0.75)$&1&	1&	1&	1&1&	1&	1&	1 &1&	1&	1&	1 \\
 \\
 \\
  \\
  \\
  \\
  \\
$\Phi^{-1}(0.9)$&1&	1&	1&	1&1&	1&	1&	1&1&	1&	1&	1
\\
\\
\\
\\
 \\
 \\
 \\
  \\
  \\
  \\
  \\
  \\
%
%
%
\mc{13}{c}{\textbf{Correlation=0.6}} \\
\\
\\
\\
$\tau$&&&RC1 &&&&RC2 &&&& RR\\
\hline
\\
\\
 & EV\_X &  IV\_X & EV\_RM & IV\_RM &  EV\_X &  IV\_X & EV\_RM & IV\_RM&  EV\_X &  IV\_X & EV\_RM & IV\_RM\\
\\
\\
$\Phi^{-1}(0.1)$ &0.936	&1	&0.908&	0.886&1&	1&	1&	0.999&0.768&	0.983&	0.790&	0.738

 \\
 \\
  \\
  \\
  \\
  \\
  $\Phi^{-1}(0.25)$&1	&1	&1&	1&1	&1	&1&	1& 0.995	&1	&0.996	&0.996
 \\
 \\
  \\
  \\
  \\
  \\
$\Phi^{-1}(0.5)$&1	&1	&1&	1&1	&1	&1&	1&1	&1	&1&	1 \\
 \\
 \\
  \\
  \\
  \\
  \\
$\Phi^{-1}(0.75)$ &1	&1	&1&	1&1	&1	&1&	1&1	&1	&1&	1 \\
 \\
 \\
  \\
  \\
  \\
  \\
$\Phi^{-1}(0.9)$ &0.999&	0.999&	0.984&	0.983&1	&1	&1&	1&1	&1	&1&	1
   \\
\\
\\
\\
 \\
 \\
 \\
  \\
  \\
  \\
  \\
  \\
%
%
%
\mc{13}{c}{\textbf{Correlation=0.4}} \\
\\
\\
\\
$\tau$&&&RC1 &&&&RC2 &&&& RR\\
\hline
\\
\\
 & EV\_X &  IV\_X & EV\_RM & IV\_RM &  EV\_X &  IV\_X & EV\_RM & IV\_RM&  EV\_X &  IV\_X & EV\_RM & IV\_RM\\
\\
\\
$\Phi^{-1}(0.1)$&0.146&	0.977&	0.300&	0.310& 0.985&	0.998&	0.915&	0.904&0.572&	0.980&	0.596&	0.567
 \\
 \\
  \\
  \\
  \\
  \\
  $\Phi^{-1}(0.25)$&1	&1&	0.875&	0.881&1&	1&	0.986&	0.986 &0.872&	1&	0.864&	0.814
 \\
 \\
  \\
  \\
  \\
  \\
$\Phi^{-1}(0.5)$&1	&1	&1	&0.999&1&	1&	0.998&	0.995&0.998&	1&	0.986&	0.981

 \\
 \\
  \\
  \\
  \\
  \\
$\Phi^{-1}(0.75)$&1	&1	&0.926&	0.928&1&	1&	0.987&	0.991&1	&1	&0.995&	0.997

 \\
 \\
  \\
  \\
  \\
  \\
$\Phi^{-1}(0.9)$& 0.376&	0.998&	0.430&	0.425&0.999&	1&	0.933&	0.938&0.989&	1&	0.969&	0.971

\end{tabular}
\end{center}
\footnotesize{$^a$ The values
in the cells are the convergence percent, i.e., percentage of replications over 1000 replications in which the estimation procedure converged. $^b$ $n=3,000$ with cumulative incidence of 0.5. $^c$ The estimates were obtained under unknown nuisance parameters which were estimated by a validation sample of size 500. }

\newpage
\normalsize
  {\bf Table 2. Pctgud$^a$, Rare Disease$^b$, Validation Sample Size=500$^c$} 
\begin{center}
\fontsize{7.5}{0.3}\selectfont

\begin{tabular}{m{1.09cm}|  m{0.74cm} m{0.74cm} m{0.74cm}  m{0.74cm}| m{0.74cm} m{0.74cm} m{0.74cm} m{0.74cm}| m{0.74cm} m{0.74cm}  m{0.74cm} m{0.74cm}| m{0.74cm} m{0.74cm}  m{0.74cm} m{0.74cm}}
\mc{13}{c}{\textbf{Correlation=0.8}} \\
\\
\\
\\
$\tau$&&&RC1 &&&&RC2 &&&& RR\\
\hline
\\
\\
 & EV\_X &  IV\_X & EV\_RM & IV\_RM &  EV\_X &  IV\_X & EV\_RM & IV\_RM&  EV\_X &  IV\_X & EV\_RM & IV\_RM\\
\\
\\
$\Phi^{-1}(0.1)$ &0.991	&0.994	&0.995&	0.989&1&	1&	1&	1&0.944&	0.950&	0.957&	0.936
 \\
 \\
  \\
  \\
  \\
  \\
  $\Phi^{-1}(0.25)$&1	&1	&1&	1&1&	1&	1&	1&1&	1&	1&	1 \\
 \\
 \\
  \\
  \\
  \\
  \\
$\Phi^{-1}(0.5)$ &1	&1	&1&	1&1&	1&	1&	1&1&	1&	1&	1
 \\
 \\
  \\
  \\
  \\
  \\
$\Phi^{-1}(0.75)$ &1	&1	&1&	1&1&	1&	1&	1&1&	1&	1&	1
 \\
 \\
  \\
  \\
  \\
  \\
$\Phi^{-1}(0.9)$&0.995&	0.988&	0.989&	0.990&1&	0.998&	0.995&	0.997&1&	1&	1&	1

 \\
 \\
  \\
  \\
  \\
  \\
%
%
\mc{13}{c}{\textbf{Correlation=0.6}} \\
\\
\\
\\
$\tau$&&&RC1 &&&&RC2 &&&& RR\\
\hline
\\
\\
 & EV\_X &  IV\_X & EV\_RM & IV\_RM &  EV\_X &  IV\_X & EV\_RM & IV\_RM&  EV\_X &  IV\_X & EV\_RM & IV\_RM\\
\\
$\Phi^{-1}(0.1)$&0.850&	0.858&	0.822&	0.810&1&	0.999&	0.999&	0.995 &0.737&	0.732&	1&	0.749
 \\
 \\
  \\
  \\
  \\
  \\
  $\Phi^{-1}(0.25)$ &1&	1&	1&	0.999&1&	1&	1&	1&0.993&	0.986&	0.992&	0.972

 \\
 \\
  \\
  \\
  \\
  \\
$\Phi^{-1}(0.5)$ &1&	1&	1&	1&1&	1&	1&	1&1&	1&	1&	1 \\
 \\
 \\
  \\
  \\
  \\
  \\
$\Phi^{-1}(0.75)$&1&	1&	1&	0.999 &1&	1&	1&	1&1&	1&	1&	1 \\
 \\
 \\
  \\
  \\
  \\
  \\
$\Phi^{-1}(0.9)$ &0.996&	0.965&	0.985&	0.982&1&	0.997&	1&	1&1&	1&	1&	1

 \\
\\
\\
\\
 \\
 \\
 \\
  \\
  \\
  \\
  \\
  \\
%
%
\mc{13}{c}{\textbf{Correlation=0.4}} \\
\\
\\
\\
$\tau$&&&RC1 &&&&RC2 &&&& RR\\
\hline
\\
\\
 & EV\_X &  IV\_X & EV\_RM & IV\_RM &  EV\_X &  IV\_X & EV\_RM & IV\_RM&  EV\_X &  IV\_X & EV\_RM & IV\_RM\\
\\
\\
$\Phi^{-1}(0.1)$ &0.118&	0.329&	0.266&	0.279&0.984&	0.888&	0.902&	0.879 &0.602&	0.513&	0.649&	0.647
 \\
 \\
  \\
  \\
  \\
  \\
  $\Phi^{-1}(0.25)$& 0.983&	0.964&	0.854&	0.841&1	&0.984&	0.991&	0.975&0.813&	0.798&	0.831&	0.806
 \\
 \\
  \\
  \\
  \\
  \\
$\Phi^{-1}(0.5)$&1	&0.999&	1	&0.997&1&	0.998&	0.998&	0.993&0.997&	0.991&	0.988&	0.968
 \\
 \\
  \\
  \\
  \\
  \\
$\Phi^{-1}(0.75)$&1&	0.990&	0.956&	0.940&1	&0.987&	0.982&	0.979&1	&1	&0.997&	0.994

 \\
 \\
  \\
  \\
  \\
  \\
$\Phi^{-1}(0.9)$&0.608&	0.867&	1&	0.532&1&	0.966&	0.928&	0.923&1&	0.998&	0.995&	0.988

\end{tabular}
\end{center}
\footnotesize{$^a$ The values
in the cells are the convergence percent, i.e., percentage of replications over 1000 replications in which the estimation procedure converged. $^b$ $n=50,000$ with cumulative incidence of 0.03. $^c$ The estimates were obtained under unknown nuisance parameters which were estimated by a validation sample of size 500.}

\normalsize
\begin{landscape}
\subsection*{B.4. \enspace Confidence Intervals}
{\bf Table 1. Confidence Interval$^a$, Corelation=0.8, Common Disease$^b$, Validation Sample Size=500$^c$}
\begin{center}
\fontsize{7.5}{0.3}\selectfont
\begin{tabular}{m{1.09cm}|  m{0.74cm} m{0.74cm} m{0.74cm}  m{0.74cm}| m{0.74cm} m{0.74cm} m{0.74cm} m{0.74cm}| m{0.74cm} m{0.74cm}  m{0.74cm} m{0.74cm}| m{0.74cm} m{0.74cm}  m{0.74cm} m{0.74cm}}
\mc{13}{c}{\textbf{Finite Sample Confidence Interval of \boldmath{$\beta$}, (\boldmath{$\beta$}, \boldmath{$\omega$})=(0.405, 0.693)}} \\
\\
\\
\\
$\tau$&&&RC1 &&&&RC2 &&&& RR1\\
\hline
\\
\\
 & EV\_X &  IV\_X & EV\_RM & IV\_RM &  EV\_X &  IV\_X & EV\_RM & IV\_RM&  EV\_X &  IV\_X & EV\_RM & IV\_RM\\
\\
\\
$\Phi^{-1}(0.1)$ &0.951&	1.000&	0.969&	0.986&0.894&	0.957&	0.929&	0.945&0.971&	0.980&	0.982&	0.971\\
 \\
 \\
  \\
  \\
  \\
  \\

  $\Phi^{-1}(0.25)$ &0.872&	0.938&	0.881&	0.891&0.885	&0.928&	0.905&	0.890&0.961&	0.979&	0.961&	0.967 \\
 \\
 \\
  \\
  \\
  \\
  \\
$\Phi^{-1}(0.5)$ &0.722&	0.802&	0.754&	0.763&0.918&	0.944&	0.926&	0.916&0.948	&0.955&	0.949&	0.958  \\
 \\
 \\
  \\
  \\
  \\
  \\
$\Phi^{-1}(0.75)$ &0.775&	0.821&	0.790&	0.865&0.948&	0.951&	0.949&	0.946&0.940	&0.941&	0.935&	0.954 \\
 \\
 \\
  \\
  \\
  \\
  \\
$\Phi^{-1}(0.9)$ &0.927	&0.924&	0.913&	0.958&0.958&	0.948&	0.962&	0.974&0.934	& 0.938&	0.942&	0.960  \\
\\
\\
\\
 \\
 \\
 \\
  \\
  \\
  \\
  \\
  \\

\mc{13}{c}{\textbf{Finite Sample Confidence Interval of \boldmath{$\omega$}, (\boldmath{$\beta$}, \boldmath{$\omega$})=(0.405, 0.693)}} \\
\\
\\
\\
$\tau$&&&RC1 &&&&RC2 &&&& RR1\\
\hline
\\
\\
 & EV\_X &  IV\_X & EV\_RM & IV\_RM &  EV\_X &  IV\_X & EV\_RM & IV\_RM&  EV\_X &  IV\_X & EV\_RM & IV\_RM\\
\\
\\
$\Phi^{-1}(0.1)$& 0.914&	1.000&	0.934&	0.957&0.842&	0.922&	0.863&	0.874&0.986	&0.988&	0.996&	0.981  \\

 \\
 \\
  \\
  \\
  \\
  \\
  $\Phi^{-1}(0.25)$ &0.607&	0.806&	0.646&	0.671&0.736&	0.826&	0.771&	0.787&0.939	& 0.939&	0.954&	0.980\\
 \\
 \\
  \\
  \\
  \\
  \\
$\Phi^{-1}(0.5)$ &0.272&	0.448&	0.278&	0.365&0.775&	0.855&	0.782&	0.834&0.824	& 0.831	&0.848&	0.905\\
 \\
 \\
  \\
  \\
  \\
  \\

$\Phi^{-1}(0.75)$ &0.464&	0.598&	0.458&	0.554&0.872&	0.902&	0.853&	0.905&0.748&	0.758&	0.757&	0.844 \\
 \\
 \\
  \\
  \\
  \\
  \\
$\Phi^{-1}(0.9)$ &0.844&	0.878&	0.834&	0.867&0.912&	0.927&	0.891&	0.927&0.800&	0.815&	0.774&	0.880\\
\end{tabular}
\end{center}
\footnotesize{$^a$ Empirical coverage rate of nominal $95\%$ confidence interval. $^b$ $n=3,000$ with cumulative incidence of 0.5. $^c$ The estimates were obtained under unknown nuisance parameters which were estimated by a validation sample of size 500.}
\end{landscape}

\begin{landscape}
{\bf Table 2. Confidence Interval$^a$, Corelation=0.6, Common Disease$^b$, Validation Sample Size=500$^c$}
\begin{center}
\fontsize{7.5}{0.3}\selectfont
\begin{tabular}{m{1.09cm}|  m{0.74cm} m{0.74cm} m{0.74cm}  m{0.74cm}| m{0.74cm} m{0.74cm} m{0.74cm} m{0.74cm}| m{0.74cm} m{0.74cm}  m{0.74cm} m{0.74cm}| m{0.74cm} m{0.74cm}  m{0.74cm} m{0.74cm}}
\mc{13}{c}{\textbf{Finite Sample Confidence Interval of \boldmath{$\beta$}, (\boldmath{$\beta$}, \boldmath{$\omega$})=(0.405, 0.693)}} \\
\\
\\
\\
$\tau$&&&RC1 &&&&RC2 &&&& RR1\\
\hline
\\
\\
 & EV\_X &  IV\_X & EV\_RM & IV\_RM &  EV\_X &  IV\_X & EV\_RM & IV\_RM&  EV\_X &  IV\_X & EV\_RM & IV\_RM\\
\\
\\
$\Phi^{-1}(0.1)$ &0.981&	1.000&	0.997&	0.996&0.905&	0.981&	0.977&	0.982&0.939	& 0.993	&0.956	&0.923\\
 \\
 \\
  \\
  \\
  \\
  \\
  $\Phi^{-1}(0.25)$ & 0.917&	0.997&	0.951&	0.953&0.890&	0.959&	0.952&	0.945&0.961&	0.986&	0.973&	0.961 \\
 \\
 \\
  \\
  \\
  \\
  \\
$\Phi^{-1}(0.5)$ & 0.653&	0.713&	0.773&	0.672&0.929&	0.944&	0.962&	0.949&0.951&	0.958&	0.964&	0.960 \\
 \\
 \\
  \\
  \\
  \\
  \\
$\Phi^{-1}(0.75)$ &0.698&	0.774&	0.831&	0.767&0.961&	0.943&	0.983&	0.980&0.947	& 0.956	&0.955	&0.971 \\
 \\
 \\
  \\
  \\
  \\
  \\
$\Phi^{-1}(0.9)$ & 0.922&	0.921&	0.957&	0.954&0.987	&0.937&	0.976&	0.991&0.948	& 0.951&	0.955&	0.964 \\
\\
\\
\\
 \\
 \\
 \\
  \\
  \\
  \\
  \\
  \\

\mc{13}{c}{\textbf{Finite Sample Confidence Interval of \boldmath{$\omega$}, (\boldmath{$\beta$}, \boldmath{$\omega$})=(0.405, 0.693)}} \\
\\
\\
\\
$\tau$&&&RC1 &&&&RC2 &&&& RR1\\
\hline
\\
\\
 & EV\_X &  IV\_X & EV\_RM & IV\_RM &  EV\_X &  IV\_X & EV\_RM & IV\_RM&  EV\_X &  IV\_X & EV\_RM & IV\_RM\\
\\
\\
$\Phi^{-1}(0.1)$&0.987	&1.000&	0.990&	0.993&0.821&	0.988&	0.861&	0.872&0.961&	0.999&	0.977&	0.944  \\

 \\
 \\
  \\
  \\
  \\
  \\
  $\Phi^{-1}(0.25)$ &0.640&	0.946&	0.684&	0.723&0.747&	0.909&	0.770&	0.765&0.989	& 0.982&	0.985&	0.984\\
 \\
 \\
  \\
  \\
  \\
  \\
$\Phi^{-1}(0.5)$ &0.134&	0.250&	0.154&	0.208&0.794	&0.903&	0.783&	0.780&0.872	& 0.757&	0.900&	0.931\\
 \\
 \\
  \\
  \\
  \\
  \\
$\Phi^{-1}(0.75)$ &0.511&	0.563&	0.514&	0.507&0.885&	0.947&	0.863&	0.875&0.805	& 0.729&	0.817&	0.888 \\
 \\
 \\
  \\
  \\
  \\
  \\
$\Phi^{-1}(0.9)$ &0.947&	0.928&	0.937&	0.940&0.932&	0.970&	0.916&	0.933&0.876&	0.850&	0.856&	0.944\\
\end{tabular}
\end{center}
\footnotesize{$^a$ Empirical coverage rate of nominal $95\%$ confidence interval. $^b$ $n=3,000$ with cumulative incidence of 0.5. $^c$ The estimates were obtained under unknown nuisance parameters which were estimated by a validation sample of size 500.}
\end{landscape}

\begin{landscape}
{\bf Table 3. Confidence Interval$^a$, Corelation=0.4, Common Disease$^b$, Validation Sample Size=500$^c$}
\begin{center}
\fontsize{7.5}{0.3}\selectfont
\begin{tabular}{m{1.09cm}|  m{0.74cm} m{0.74cm} m{0.74cm}  m{0.74cm}| m{0.74cm} m{0.74cm} m{0.74cm} m{0.74cm}| m{0.74cm} m{0.74cm}  m{0.74cm} m{0.74cm}| m{0.74cm} m{0.74cm}  m{0.74cm} m{0.74cm}}
\mc{13}{c}{\textbf{Finite Sample Confidence Interval of \boldmath{$\beta$}, (\boldmath{$\beta$}, \boldmath{$\omega$})=(0.405, 0.693)}} \\
\\
\\
\\
$\tau$&&&RC1 &&&&RC2 &&&& RR1\\
\hline
\\
\\
 & EV\_X &  IV\_X & EV\_RM & IV\_RM &  EV\_X &  IV\_X & EV\_RM & IV\_RM&  EV\_X &  IV\_X & EV\_RM & IV\_RM\\
\\
\\
$\Phi^{-1}(0.1)$ &1.000	&1.000&	1.000	&0.994&0.965&	0.963&	1.000&	1.000&0.797&	0.997&	0.884&	0.802\\
 \\
 \\
  \\
  \\
  \\
  \\
  $\Phi^{-1}(0.25)$ &0.987	&1.000	&1.000&	0.998&0.947&	0.943&	1.000&	1.000&0.901	& 0.994&	0.942&	0.925  \\
 \\
 \\
  \\
  \\
  \\
  \\
$\Phi^{-1}(0.5)$ & 0.752&	0.802&	0.998&	0.988&0.965&	0.937&	1.000&	1.000&0.941	& 0.975	&0.944&	0.961 \\
 \\
 \\
  \\
  \\
  \\
  \\
$\Phi^{-1}(0.75)$ &0.755&	0.869&	0.990&	0.991&0.994&	0.934&	0.995&	0.999&0.950	& 0.960&	0.971&	0.978 \\
 \\
 \\
  \\
  \\
  \\
  \\
$\Phi^{-1}(0.9)$ &0.957&	0.947&	0.919&	0.965&1.000&	0.947&	0.985&	0.992&0.955	& 0.970&	0.960&	0.973 \\
\\
\\
\\
 \\
 \\
 \\
  \\
  \\
  \\
  \\
  \\

\mc{13}{c}{\textbf{Finite Sample Confidence Interval of \boldmath{$\omega$}, (\boldmath{$\beta$}, \boldmath{$\omega$})=(0.405, 0.693)}} \\
\\
\\
\\
$\tau$&&&RC1 &&&&RC2 &&&& RR1\\
\hline
\\
\\
 & EV\_X &  IV\_X & EV\_RM & IV\_RM &  EV\_X &  IV\_X & EV\_RM & IV\_RM&  EV\_X &  IV\_X & EV\_RM & IV\_RM\\
\\
\\
$\Phi^{-1}(0.1)$& 1.000&	1.000&	0.993&	0.997&0.919&	0.981&	0.973&	0.966&0.851	& 0.998&	0.916&	0.850 \\

 \\
 \\
  \\
  \\
  \\
  \\
  $\Phi^{-1}(0.25)$ &0.939&	1.000&	0.909&	0.913&0.879	&0.972&	0.925&	0.916&0.956	& 0.998&	0.964&	0.942\\
 \\
 \\
  \\
  \\
  \\
  \\
$\Phi^{-1}(0.5)$ &0.261&	0.282&	0.364&	0.377&0.889&	0.955&	0.909&	0.904&0.989&	0.803&	0.967&	0.982\\
 \\
 \\
  \\
  \\
  \\
  \\

$\Phi^{-1}(0.75)$ & 0.878&	0.698&	0.826&	0.803&0.927&	0.968&	0.949&	0.941&0.975	& 0.803&	0.961&	0.986\\
 \\
 \\
  \\
  \\
  \\
  \\
$\Phi^{-1}(0.9)$ &1.000&	0.951&	0.991&	0.986&0.952&	0.984&	0.980&	0.979&0.990&	0.912&	0.979&	0.987\\
\end{tabular}
\end{center}
\footnotesize{$^a$ Empirical coverage rate of nominal $95\%$ confidence interval. $^b$ $n=3,000$ with cumulative incidence of 0.5. $^c$ The estimates were obtained under unknown nuisance parameters which were estimated by a validation sample of size 500.}
\end{landscape}

\begin{landscape}
{\bf Table 4. Confidence Interval$^a$, Corelation=0.8, Rare Disease$^b$, Validation Sample Size=500$^c$}
\begin{center}
\fontsize{7.5}{0.3}\selectfont
\begin{tabular}{m{1.09cm}|  m{0.74cm} m{0.74cm} m{0.74cm}  m{0.74cm}| m{0.74cm} m{0.74cm} m{0.74cm} m{0.74cm}| m{0.74cm} m{0.74cm}  m{0.74cm} m{0.74cm}| m{0.74cm} m{0.74cm}  m{0.74cm} m{0.74cm}}
\mc{13}{c}{\textbf{Finite Sample Confidence Interval of \boldmath{$\beta$}, (\boldmath{$\beta$}, \boldmath{$\omega$})=(0.405, 0.693)}} \\
\\
\\
\\
$\tau$&&&RC1 &&&&RC2 &&&& RR1\\
\hline
\\
\\
 & EV\_X &  IV\_X & EV\_RM & IV\_RM &  EV\_X &  IV\_X & EV\_RM & IV\_RM&  EV\_X &  IV\_X & EV\_RM & IV\_RM\\
\\
\\
$\Phi^{-1}(0.1)$ &0.967&	0.993&	0.980&	0.985&0.939&	0.941&	0.952&	0.957&0.939	& 0.948&	0.942&	0.940\\
 \\
 \\
  \\
  \\
  \\
  \\
  $\Phi^{-1}(0.25)$ & 0.889&	0.888&	0.900&	0.894&0.861&	0.868&	0.879&	0.867&0.963&	0.967&	0.966&	0.964 \\
 \\
 \\
  \\
  \\
  \\
  \\
$\Phi^{-1}(0.5)$ & 0.534&	0.527&	0.547&	0.604&0.759&	0.760&	0.800&	0.770&0.956	& 0.955	&0.963	&0.968 \\
 \\
 \\
  \\
  \\
  \\
  \\
$\Phi^{-1}(0.75)$ &0.327&	0.342&	0.367&	0.600&0.822	&0.826&	0.842&	0.801&0.946	& 0.948&	0.957&	0.964 \\
 \\
 \\
  \\
  \\
  \\
  \\
$\Phi^{-1}(0.9)$ &0.511&	0.535&	0.449&	0.827&0.893&	0.903&	0.924&	0.905&0.947	& 0.943&	0.957&	0.963  \\
\\
\\
\\
 \\
 \\
 \\
  \\
  \\
  \\
  \\
  \\

\mc{13}{c}{\textbf{Finite Sample Confidence Interval of \boldmath{$\omega$}, (\boldmath{$\beta$}, \boldmath{$\omega$})=(0.405, 0.693)}} \\
\\
\\
\\
$\tau$&&&RC1 &&&&RC2 &&&& RR1\\
\hline
\\
\\
 & EV\_X &  IV\_X & EV\_RM & IV\_RM &  EV\_X &  IV\_X & EV\_RM & IV\_RM&  EV\_X &  IV\_X & EV\_RM & IV\_RM\\
\\
\\
$\Phi^{-1}(0.1)$&0.972&	0.994	&0.978&	0.996&0.935&	0.941&	0.941&	0.933&0.942	& 0.951	&0.945&	0.942   \\

 \\
 \\
  \\
  \\
  \\
  \\
  $\Phi^{-1}(0.25)$ &0.865&	0.867&	0.868&	0.856&0.873&	0.871&	0.880&	0.859&0.968	& 0.965&	0.971&	0.968\\
 \\
 \\
  \\
  \\
  \\
  \\
$\Phi^{-1}(0.5)$ &0.579&	0.549&	0.573&	0.589&0.866&	0.864&	0.896&	0.892&0.955&	0.942&	0.956&	0.956\\
 \\
 \\
  \\
  \\
  \\
  \\

$\Phi^{-1}(0.75)$ &0.662&	0.635&	0.663&	0.672&0.959&	0.938&	0.966&	0.972&0.952	& 0.925&0.963&	0.962 \\
 \\
 \\
  \\
  \\
  \\
  \\
$\Phi^{-1}(0.9)$ &0.947&0.921&	0.457&	0.943&0.929&	0.926&	0.945&	0.980&0.956	& 0.936	&0.964&	0.963\\
\end{tabular}
\end{center}
\footnotesize{$^a$ Empirical coverage rate of nominal $95\%$ confidence interval. $^b$ $n=50,000$ with cumulative incidence of 0.03. $^c$ The estimates were obtained under unknown nuisance parameters which were estimated by a validation sample of size 500.}
\end{landscape}

\begin{landscape}
{\bf Table 5. Confidence Interval$^a$, Corelation=0.6, Rare Disease$^b$, Validation Sample Size=500$^c$}
\begin{center}
\fontsize{7.5}{0.3}\selectfont
\begin{tabular}{m{1.09cm}|  m{0.74cm} m{0.74cm} m{0.74cm}  m{0.74cm}| m{0.74cm} m{0.74cm} m{0.74cm} m{0.74cm}| m{0.74cm} m{0.74cm}  m{0.74cm} m{0.74cm}| m{0.74cm} m{0.74cm}  m{0.74cm} m{0.74cm}}
\mc{13}{c}{\textbf{Finite Sample Confidence Interval of \boldmath{$\beta$}, (\boldmath{$\beta$}, \boldmath{$\omega$})=(0.405, 0.693)}} \\
\\
\\
\\
$\tau$&&&RC1 &&&&RC2 &&&& RR1\\
\hline
\\
\\
 & EV\_X &  IV\_X & EV\_RM & IV\_RM &  EV\_X &  IV\_X & EV\_RM & IV\_RM&  EV\_X &  IV\_X & EV\_RM & IV\_RM\\
\\
\\
$\Phi^{-1}(0.1)$ &0.959&	0.995&	0.992&	0.993&0.926&	0.939&	0.988&	0.984&0.863	& 0.892	&0.880&	0.885\\
 \\
 \\
  \\
  \\
  \\
  \\
  $\Phi^{-1}(0.25)$ & 0.893&	0.897&	0.942&	0.940&0.813&	0.800&	0.870&	0.863&0.946&	0.952&	0.952&	0.955 \\
 \\
 \\
  \\
  \\
  \\
  \\
$\Phi^{-1}(0.5)$ & 0.273&	0.247&	0.359&	0.385&0.721&	0.671&	0.770&	0.754&0.963	& 0.965&	0.971&	0.979 \\
 \\
 \\
  \\
  \\
  \\
  \\
$\Phi^{-1}(0.75)$ &0.074&	0.085&	0.278&	0.343&0.831&0.737&	0.839&	0.809&0.966	& 0.966&	0.982&	0.963 \\
 \\
 \\
  \\
  \\
  \\
  \\
$\Phi^{-1}(0.9)$ &0.318&	0.367&	0.859&	0.697&0.923&	0.863&	0.940&	0.917&0.947&	0.958&	0.976&	0.951  \\
\\
\\
\\
 \\
 \\
 \\
  \\
  \\
  \\
  \\
  \\

\mc{13}{c}{\textbf{Finite Sample Confidence Interval of \boldmath{$\omega$}, (\boldmath{$\beta$}, \boldmath{$\omega$})=(0.405, 0.693)}} \\
\\
\\
\\
$\tau$&&&RC1 &&&&RC2 &&&& RR1\\
\hline
\\
\\
 & EV\_X &  IV\_X & EV\_RM & IV\_RM &  EV\_X &  IV\_X & EV\_RM & IV\_RM&  EV\_X &  IV\_X & EV\_RM & IV\_RM\\
\\
\\
$\Phi^{-1}(0.1)$&0.966&	0.998&	0.993	&0.995&0.902&0.923&	0.942&	0.915&0.875	& 0.900&	0.894&	0.887  \\

 \\
 \\
  \\
  \\
  \\
  \\
  $\Phi^{-1}(0.25)$ &0.862&0.862&	0.859&	0.846&0.826&	0.824&	0.833&	0.814&0.956&	0.960&	0.949&	0.960\\
 \\
 \\
  \\
  \\
  \\
  \\
$\Phi^{-1}(0.5)$ &0.313&	0.249&	0.347&	0.342&0.880&	0.837&	0.893&	0.870&0.960	& 0.968&0.966	&0.968\\
 \\
 \\
  \\
  \\
  \\
  \\

$\Phi^{-1}(0.75)$ &0.593&	0.476&	0.605&	0.599&0.964	&0.950&	0.980&	0.970&0.961	& 0.951&	0.973&	0.949 \\
 \\
 \\
  \\
  \\
  \\
  \\
$\Phi^{-1}(0.9)$ &0.973&	0.940&	0.751&	0.976&0.930&	0.931&	0.988&	0.986&0.955	& 0.953&	0.971&	0.958\\
\end{tabular}
\end{center}
\footnotesize{$^a$ Empirical coverage rate of nominal $95\%$ confidence interval. $^b$ $n=50,000$ with cumulative incidence of 0.03. $^c$ The estimates were obtained under unknown nuisance parameters which were estimated by a validation sample of size 500.}
\end{landscape}

\begin{landscape}
{\bf Table 6. Confidence Interval$^a$, Corelation=0.4, Rare Disease$^b$, Validation Sample Size=500$^c$}
\begin{center}
\fontsize{7.5}{0.3}\selectfont
\begin{tabular}{m{1.09cm}|  m{0.74cm} m{0.74cm} m{0.74cm}  m{0.74cm}| m{0.74cm} m{0.74cm} m{0.74cm} m{0.74cm}| m{0.74cm} m{0.74cm}  m{0.74cm} m{0.74cm}| m{0.74cm} m{0.74cm}  m{0.74cm} m{0.74cm}}
\mc{13}{c}{\textbf{Finite Sample Confidence Interval of \boldmath{$\beta$}, (\boldmath{$\beta$}, \boldmath{$\omega$})=(0.405, 0.693)}} \\
\\
\\
\\
$\tau$&&&RC1 &&&&RC2 &&&& RR1\\
\hline
\\
\\
 & EV\_X &  IV\_X & EV\_RM & IV\_RM &  EV\_X &  IV\_X & EV\_RM & IV\_RM&  EV\_X &  IV\_X & EV\_RM & IV\_RM\\
\\
\\
$\Phi^{-1}(0.1)$ &0.992&	0.985&	1.000&	1.000&0.968&	0.981&	1.000&	1.000&0.764	& 0.798&	0.808	&0.759\\
 \\
 \\
  \\
  \\
  \\
  \\
  $\Phi^{-1}(0.25)$ &0.991&	0.995&	1.000&	0.999&0.911&	0.920&	1.000&	1.000&0.899	&0.919	&0.917&	0.898  \\
 \\
 \\
  \\
  \\
  \\
  \\
$\Phi^{-1}(0.5)$ & 0.302&	0.496&	0.958&	0.914&0.893&	0.853&	1.000&	1.000&0.947&	0.960&	0.955&	0.971 \\
 \\
 \\
  \\
  \\
  \\
  \\
$\Phi^{-1}(0.75)$ &0.069&	0.372&	0.989&	0.956&0.938&	0.800&	0.997&	0.994&0.977&	0.969&	0.991	&0.975 \\
 \\
 \\
  \\
  \\
  \\
  \\
$\Phi^{-1}(0.9)$ &0.490&	0.809&	0.984&	0.993&0.987&	0.893&	0.993&	0.987&0.969&	0.965&	0.985&	0.972 \\
\\
\\
\\
 \\
 \\
 \\
  \\
  \\
  \\
  \\
  \\

\mc{13}{c}{\textbf{Finite Sample Confidence Interval of \boldmath{$\omega$}, (\boldmath{$\beta$}, \boldmath{$\omega$})=(0.405, 0.693)}} \\
\\
\\
\\
$\tau$&&&RC1 &&&&RC2 &&&& RR1\\
\hline
\\
\\
 & EV\_X &  IV\_X & EV\_RM & IV\_RM &  EV\_X &  IV\_X & EV\_RM & IV\_RM&  EV\_X &  IV\_X & EV\_RM & IV\_RM\\
\\
\\
$\Phi^{-1}(0.1)$&  0.992&0.988&	1.000&	1.000&0.956&	0.973&	0.989&	0.985&0.777	& 0.809&	0.814&	0.777 \\

 \\
 \\
  \\
  \\
  \\
  \\
  $\Phi^{-1}(0.25)$ &0.985&	0.988&	0.945&	0.939&0.911&	0.834&	0.946&	0.936&0.905	& 0.932&	0.897&	0.901\\
 \\
 \\
  \\
  \\
  \\
  \\
$\Phi^{-1}(0.5)$ &0.363&	0.250&	0.444&	0.431&0.938&	0.794&	0.950&	0.934&0.963&	0.970&	0.944&	0.960\\
 \\
 \\
  \\
  \\
  \\
  \\

$\Phi^{-1}(0.75)$ &0.901&	0.677&	0.850&	0.817&0.968	&0.883&	0.986&	0.981&0.974	& 0.979&	0.987&	0.967 \\
 \\
 \\
  \\
  \\
  \\
  \\
$\Phi^{-1}(0.9)$ &0.995&	0.974&	0.945&	0.991&0.955&	0.951&	0.997&	0.992&0.965	& 0.973&	0.989&	0.969\\
\end{tabular}
\end{center}
\footnotesize{$^a$ Empirical coverage rate of nominal $95\%$ confidence interval. $^b$ $n=50,000$ with cumulative incidence of 0.03. $^c$ The estimates were obtained under unknown nuisance parameters which were estimated by a validation sample of size 500.}
\end{landscape}

\normalsize

\bibliographystyle{acm}
\nocite{*}
\bibliography{refer_valid1}

\end{document}